\documentclass[dvips,12pt]{article}
\usepackage{hyperref}
\usepackage{amsfonts}
\usepackage{amsmath, amssymb}
\usepackage{graphicx}
\usepackage{psfrag}
\usepackage{youngtab}
\Yboxdim{5pt}

\newcommand{\ra}{\rightarrow}

\newcommand{\be}{\begin{equation}}
\newcommand{\ee}{\end{equation}}
\newcommand{\ba}{\begin{eqnarray}}
\newcommand{\ea}{\end{eqnarray}}
\newcommand{\bi}{\begin{itemize}}
\newcommand{\ei}{\end{itemize}}

\newcommand{\Tr}{{\rm Tr}}

\newcommand{\Z}{{\bf Z}}
\newcommand{\R}{{\rm R}}

\newcommand{\p}{\partial}

\newcommand{\Ncal}{{\mathcal N}}

\newcommand{\Jcal}{{\mathcal J}}
\newcommand{\Acal}{{\mathcal A}}

\newcommand{\Kahler}{K\"{a}hler }

\newcommand{\nn}{\nonumber}
\newcommand{\mo}{{-1}} 

\newcommand{\f}{\frac}
\newcommand{\half}{\frac{1}{2}}
\newcommand{\oo}{\frac{1}}

\def\Dslash{\,\,{\raise.15ex\hbox{/}\mkern-12mu D}}
\def\Dbarslash{\,\,{\raise.15ex\hbox{/}\mkern-12mu {\bar D}}}
\def\delslash{\,\,{\raise.15ex\hbox{/}\mkern-9mu \partial}}
\def\delbarslash{\,\,{\raise.15ex\hbox{/}\mkern-9mu {\bar\partial}}}
\def\pslash{\,\,{\raise.15ex\hbox{/}\mkern-9mu p}}
\def\calDslash{\,\,{\raise.15ex\hbox{/}\mkern-12mu {\cal D}}}

\renewcommand{\bar}{\overline}

\renewcommand{\hat}{\widehat}

\begin{document}
\baselineskip=15.5pt
\renewcommand{\theequation}{\arabic{section}.\arabic{equation}}
\pagestyle{plain} \setcounter{page}{1}
\bibliographystyle{utcaps}
\begin{titlepage}

\rightline{\small{\tt CPHT-RR022.0407}}
\rightline{\small{\tt NSF-KITP-07-106}}
\rightline{\tt arXiv:0704.3080}
\begin{center}

\vskip 3 cm
\centerline{{\Large {\bf $D$-branes as  a Bubbling Calabi-Yau}}}
\vskip 1.5cm
{\large Jaume Gomis}\footnote{\tt jgomis at perimeterinstitute.ca}
{\large and Takuya Okuda}\footnote{\tt takuya at kitp.ucsb.edu}

\vskip 1cm

Perimeter Institute for Theoretical Physics

Waterloo, Ontario N2L 2Y5, Canada$^{1}$

and

Kavli Institute for Theoretical Physics

University of California,
Santa Barbara

CA 93106, USA${}^2$

\vskip 1.5cm

{\bf Abstract}

\end{center}

We prove that  the open topological string partition function on  a
  $D$-brane configuration in a  Calabi-Yau manifold $X$
   takes the form of a closed topological string
partition function on a different Calabi-Yau manifold $X_b$.
This identification shows that the physics of $D$-branes in an arbitrary background $X$
of topological  string theory  can be   described  either by open+closed
 string theory in $X$ or by closed string theory in   $X_b$.
The physical interpretation of the ``bubbling" Calabi-Yau $X_b$ is  as the space obtained by letting the $D$-branes in $X$
undergo a geometric transition.
This implies, in particular, that the
 partition function of {\it closed} topological string theory on certain bubbling Calabi-Yau manifolds
 are invariants of knots in the three-sphere.


\end{titlepage}

\newpage

\tableofcontents
\section{Introduction and conclusion}

$D$-branes in a given vacuum of string theory have two alternative descriptions; either in terms of open strings or in terms of closed strings. This basic observation motivates the existence of a duality between open+closed string theory in the given vacuum
and closed string theory in the vacuum where the $D$-branes have been replaced by a non-trivial geometry with fluxes\footnote{This type of duality is
often studied in a low energy, decoupling limit where
the open+closed string theory on one side of the duality reduces to a gauge theory. Taking the same limit on the purely closed string side of the duality  in effect replaces the asymptotic geometry of the original vacuum by a new asymptotic geometry. The AdS/CFT correspondence is the prototypical example of such a ``low energy"  open/closed duality.}.

In this paper we give a very concrete realization of open/closed duality. We find an explicit relation between the partition function of open+closed topological string theory in a given Calabi-Yau $X$ and the partition function of closed topological string theory in another ``bubbling" Calabi-Yau  $X_b$:
\ba
Z_{o+c}(X)=Z_c(X_b).
\label{partequal}
\ea
The physical interpretation of $X_b$ is as the background obtained by replacing the $D$-branes in $X$ by ``fluxes" when the $D$-branes undergo a geometric transition. This equality  shows that the physics of $D$-branes in an arbitrary background $X$
of topological  string theory  can be   described  either by open+closed
 string theory in $X$ or by closed string theory in   $X_b$.

The identification of the open+closed partition function in $X$ with the closed string partition function in $X_b$ does not rely on knowing explicitly the exact answer for the partition functions, which is why the result applies in great generality. The result relies on being able to write the open string partition function in terms of   the
open Gopakumar-Vafa (GV) invariants \cite{Ooguri:1999bv,Labastida:2000yw}  and the closed string partition function in terms of the closed Gopakumar-Vafa (GV) invariants \cite{Gopakumar:1998ki}.
As reviewed in section $2$, such  a parametrization of the open string partition function is possible whenever the world-volume geometry of the $D$-branes defining the open string theory has a non-trivial  first Betti number   $b_1(L)$, where $L$ is the cycle that the $D$-branes wrap. It is for such open string theories that we can explicitly show that they are completely equivalent to a closed string theory on a ``bubbling" Calabi-Yau  space $X_b$.

In order to completely determine the open string partition function  in a Calabi-Yau $X$ we must  supply the open GV invariants in $X$ {\it and} the holonomy of the gauge field on the branes. Since the holonomy of the gauge field  encodes\footnote{The position is defined only up to Hamiltonian deformations, which are gauge symmetries of the A-model open string field theory.}
the ``position" of the branes, the open string amplitude depends on the holonomy. Following  \cite{Gomis:2006mv},
we encode the data about the holonomy matrix   in a Young tableau\footnote{See section $3$ for details.}, labeled by $R$.
Given this data we prove   that the open+closed string partition function on $X$ can be
rewritten precisely as a closed  string partition function  on another Calabi-Yau $X_b$.
Namely, the open string partition function in $X$ can be written as a closed string instanton expansion
on $X_b$, which is what the closed string partition function in topological string theory computes.

We find  an explicit formula relating the closed GV invariants in $X_b$
to the
open+closed
 GV  invariants in $X$  and the holonomy of the gauge field living on the $D$-branes. As we recall in section $2$ the GV invariants are a collection of {\it integers} in terms of which  the topological string theory partition function on a  Calabi-Yau manifold can be written down to all orders in perturbation theory. The formula we find takes  the integer open and closed GV invariants in $X$ together with the holonomy of the gauge field labeled by the Young tableau $R$ and relates them to a new set of integers,
 which
 are precisely the
 closed GV invariants in another space $X_b$!

 By using the relation we obtain between   the closed GV invariants in $X_b$ and
 the open+closed GV invariants  in  $X$ combined with the holonomy of the gauge field,
 we can explicitly identify the closed string partition function in
 $X_b$ with the open+closed string partition function in $X$.
 This computation demonstrates that the physics of $D$-branes in $X$ is completely equivalent to
 {\it closed} string physics in $X_b$. This gives a way to explicitly construct open/closed dualities
 even when the explicit expressions for the partition functions are not known.
    It  allows us to relate open string theory in $X$ with closed string theory in $X_b$.

  The topology of $X_b$ depends on the topology of $X$ and on the shape of the Young tableau $R$. If we parametrize the Young tableau by using the following coordinates\footnote{Informally, $l_{odd}$ is the number of rows
in the tableau with the same number of boxes while $l_{even}$ is the number of columns in the tableau with the same number of boxes.}
\begin{figure}[htbb]
\begin{center}
\psfrag{N}{$N$}
\psfrag{l1}{$l_1$}
\psfrag{l2}{$l_2$}
\psfrag{l2m-1}{$l_{2m-1}$}
\psfrag{l2m}{$l_{2m}$}
\includegraphics[width=60mm]{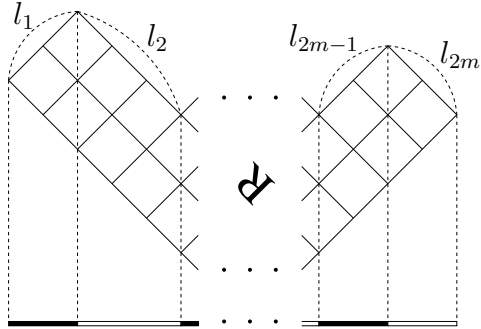}
\caption{The Young tableau $R$, shown rotated, is specified by the lengths $l_I$ of all the
edges.
Equivalently, $l_I$ denote the length of the black and white regions in
the Maya diagram.
}
\label{table}
\end{center}
\end{figure}

\noindent
then we find that $b_2(X_b)=b_2(X)+2m$, where $b_2$ is the second Betti number of the manifold. The size of the extra $2m$ two-cycles created by   replacing the branes by ``flux" is given by $t_I=g_s l_I$, with $I=1,\ldots,2m$, where $l_I$ are the coordinates of the Young tableau in Figure \ref{table}. The appearance of the extra cycles has a simple physical intepretation. The branes in $X$ can  undergo a geometric transition and be replaced by fluxes.
Fluxes in topological string theory correspond precisely to non-trivial periods of the complexified \Kahler form. In this picture, the original branes disappear and leave behind a collection of non-contractible cycles on which their flux is supported.
Therefore, the  Calabi-Yau $X_b$ captures the backreacted geometry produced by the $D$-branes in $X$. It is this picture that warrants the description of $X_b$ as a bubbling Calabi-Yau.

An interesting application of these results is to  knot invariants in
$S^3$. On the one hand, knot invariants in  $S^3$ are captured by  the
expectation value of Wilson loops in Chern-Simons theory in $S^3$ \cite{Witten:1988hf}. On the other hand, as shown in \cite{Gomis:2006mv}, a Wilson loop operator in $U(N)$ Chern-Simons theory on $S^3$ -- which is labeled by a representation $R$ and a knot $\alpha$ --  is described by a configuration of  $D$-branes or anti-branes in the resolved conifold geometry (see  \cite{Gomis:2006mv} for the details of the brane and anti-brane configuration).

Since we can now relate the open+closed GV invariants of   a brane configuration in the resolved conifold to the closed GV invariants in
    $X_b$, we arrive at the representation of knot invariants in terms of closed GV invariants in  $X_b$. This relation was already established  in
\cite{Gomis:2006mv} for the case of the unknot and for arbitrary representation $R$, where it was shown that these knot invariants are captured by the {\it closed} topological string partition function on certain bubbling Calabi-Yau manifolds.
Therefore, as
 a corollary of the results in this paper and those in \cite{Gomis:2006mv} we find a novel representation of knot invariants  for arbitrary knots in $S^3$ in terms of closed GV invariants of bubbling Calabi-Yau manifolds $X_b$!

An interesting recent development in the application of topological strings to
knot theory is the so-called categorification program \cite{Gukov:2004hz,Gukov:2005qp}.
The idea is to use the BPS Hilbert space associated with open strings on the branes realizing knots
to define more refined invariants than knot polynomials.
Our proposal in \cite{Gomis:2006mv} and in this paper is that these branes
can undergo a geometric transition to bubbling Calabi-Yau manifolds.
We are then   tempted to contemplate that the BPS Hilbert space associated
with {\it closed} strings on the bubbling Calabi-Yau manifolds
could be used define new knot invariants.

The results in  this paper confirm the expectation that
whenever we have many branes in a given open+closed string theory,
we have a dual description in terms of pure closed string theory in the backreacted geometry, where branes are replaced by non-trivial geometry with fluxes.
It would be very interesting to extend the ideas in this paper
to physical string theory.
Learning how to rewrite  open string theory in a given background as
a closed string theory in a different background would be tantamount to deriving open/closed
dualities in the physical theory.

This paper focuses on geometric transitions, namely on transitions of $D$-branes
into pure geometry with flux.
Another interesting phenomenon found in the study of Wilson loops
in $\Ncal=4$ Yang-Mills and Chern-Simons theory is that
fundamental strings describing Wilson loops can puff up into
$D$-branes.
Just like for geometric transitions one may expect
that the transition between strings and $D$-branes occurs
more generally.
The forthcoming paper \cite{Okuda:2007ai} will discuss a large class
of such transitions in the topological string setting.

The plan for the rest of the paper is as follows.
In section \ref{GVreview} we give a brief summary of the physical origin of open and closed GV invariants
and how they characterize the topological string partition function for open and closed strings.
In section \ref{open2closedGV-section} we show that the partition function of open+closed string theory in a Calabi-Yau $X$ is equal to the closed string partition function in a bubbling Calabi-Yau $X_b$. We argue that $X_b$ is the space obtained by letting the $D$-branes in $X$ undergo a geometric transition.
In section \ref{toricCY} we study the geometric transitions proposed in this paper in the context of toric Calabi-Yau manifolds and show that the transitions we propose can be
explicitly exhibited. The appendices contain the derivation of various formulas appearing in the main text.



\section{GV invariants in a nutshell}\label{GVreview}

The topological string partition function in $X$  computes certain F-terms \cite{Antoniadis:1993ze,Bershadsky:1993cx,Ooguri:1999bv} in the   effective action obtained by compactifying ten dimensional string theory on $X$. The physical origin  of GV invariants stems from the observation in \cite{Gopakumar:1998ki,Ooguri:1999bv,Labastida:2000yw} that these higher derivative terms in Type IIA string theory do not depend on the string coupling constant, and  can also be computed using an index that counts the BPS spectrum of wrapped membranes in an M-theory compactification on $X$.

 The upshot is that the topological string amplitudes exhibit  hitherto unknown integrality properties. Remarkably, the partition function can be computed to all orders in  perturbation theory in terms of the integral  invariants \cite{Gopakumar:1998ki,Ooguri:1999bv,Labastida:2000yw} associated to a given Calabi-Yau.

\medskip\medskip
\noindent{\it Closed GV invariants}
\medskip\medskip

The closed string partition function $Z_c$ on $X$ computes the supersymmetric completion of the following higher derivative term in the four dimensional effective action\footnote{In writing this term we have already turned on a  graviphoton field strength  background $F=g_s$, where $g_s$ is the topological string coupling constant. $R_+$ is the self-dual part of the curvature.}
\ba
F(g_s,t) R_+^2,
\label{higher}
\ea
where:
\ba
F(g_s,t)=\sum_{g=0}^\infty F_g(t) g_s^{2g-2}\qquad \hbox{and} \qquad Z_c(g_s,t)=\exp(F(g_s,t)).
\ea
$F_g(g_s,t)$ is the genus $g$ topological string free energy
and $g_s$ is the topological string coupling constant. The complex scalar fields ${\vec t}\equiv (t_1,\ldots,t_{b_2(X)})$ in the physical theory parametrize  the ``size" of the various two cycles in $X$
\ba
t_a=\int_{\Sigma_a}{\cal J},
\ea
where $\Sigma_a$ are an integral  basis of $H_2(X,\Z)$ and ${\cal J}$ is the complexified \Kahler form.

It has been argued by Gopakumar and Vafa \cite{Gopakumar:1998ii,Gopakumar:1998jq} that $F(g_s,t)$  can be computed in terms of integer invariants $n_g^{\vec{Q}}\in \Z$, where  $g \in \Z_{\geq 0}$  and $\vec{Q}\equiv (Q_1,Q_2,\ldots,Q_{b_2(X)})\in \Z^{b_2(X)}$. These integers $n_g^{\vec{Q}}$ are called invariant because they do not change under smooth complex structure deformations of $X$; they define an index. Roughly speaking, $n_g^{\vec{Q}}$ counts\footnote{$g$ encodes the
quantum number under   $SU(2)_L$, a  subgroup of the rotation group in the  four non-compact directions.} the number of BPS multiplets arising from membranes wrapping the class ${\vec \Sigma}\cdot{\vec Q}\in H_2(X,\Z)$. As shown in \cite{Gopakumar:1998ii,Gopakumar:1998jq} a one-loop
diagram with membranes running in the loop precisely generates the  term (\ref{higher}) in the four dimensional effective action.
By comparing the one-loop diagram with (\ref{higher}) one finds that \cite{Gopakumar:1998ii,Gopakumar:1998jq}:
\ba
Z_c(g_s,t)=M(q)^{\f{\chi(X)} 2}\cdot \exp\left(\sum_{g=0}^\infty\sum_{n=1}^\infty {\oo n\: [n]^{2g-2}} \sum_{\vec{Q}} n_g^{\vec{Q}} e^{-n\vec{Q}\cdot \vec{t}}\right).
\label{closedGV}
\ea
$[n]\equiv q^{n/2}-q^{-n/2}$ is a $q$-number, where  $q\equiv e^{-g_s}$ and
$\chi(X)$ is the Euler characteristic\footnote{For a compact Calabi-Yau manifold, $\chi(X)/2$ is the number of \Kahler moduli
minus the number of complex structure moduli.} of $X$.
The function
\ba
M(q)=\prod_{m=1}^\infty{\oo{ (1-q^m)^m}}
\label{mac}
\ea
is the MacMahon function, and arises from the contribution of $D0$-branes -- or eleven dimensional momentum -- running in the loop. From the world-sheet point of view,
this is the  contribution from constant maps from the world-sheet to $X$
  \cite{Gopakumar:1998ii,faber}.

Knowledge of the closed GV invariants $n_g^{\vec{Q}}$ in $X$ determines using (\ref{closedGV}) the closed topological string partition function in $X$ to all orders  in perturbation theory.

\medskip
\medskip
\noindent{\it Open GV invariants}
\medskip
\medskip

The open string partition function $Z_o$ in $X$ computes the supersymmetric completion of the following  term
 in the two dimensional effective action that arises by wrapping $P$ $D4$-branes on a
special Lagrangian submanifold\footnote{In order not to clutter the formulas and obscure the physics,
we will assume that $b_1(L)=1$ in writing the formulas. It is straightforward to write the corresponding formulas for $b_1(L)\geq1$.} $L\subset X$
\ba
F(g_s,t,V) R_+,
\label{higheropen}
\ea
where:
\ba
F(g_s,t,V)=\sum_{g=0}^\infty\sum_{h=1}^\infty F_{g,h}(t,V) g_s^{2g-2+h}~~\hbox{and} ~~ Z_o(g_s,t,V)=\exp(F(g_s,t,V)).
\ea
$F_{g,h}(g_s,t,V)$ is the  topological string free energy on a genus $g$ Riemann surface with $h$ boundaries, with the boundary conditions
specified by a Lagrangian submanifold $L$, which gives rise to BRST-invariant boundary conditions. $V$ is the $U(P)$ holonomy matrix that arises by integrating the gauge field on the $D4$-branes along the generator of $H_1(L,\Z)$. It corresponds to a complex scalar\footnote{We recall that the gauge group in topological string theory is complex.} field in the effective two dimensional theory living on the $D4$-branes.

It was shown in \cite{Ooguri:1999bv,Labastida:2000yw} that these terms also arise at one-loop
by integrating out BPS states that end on the $D4$-branes.
By comparing the one-loop computation with (\ref{higheropen}) one arrives at the following expression \cite{Ooguri:1999bv,Labastida:2000yw}:
\ba
Z_o(g_s,t,V)=\exp\left(\sum_{n=1}^\infty\sum_{\vec k}{\oo n} \oo{ z_{\vec k}} f_{\vec k}(q^n,e^{-n\vec{Q}\cdot \vec{t}})\hbox{Tr}_{\vec k}V^n \right).
\label{openGV}
\ea
In the computation the symmetric group $S_k$ plays a prominent role.
${\vec k}=(k_1,k_2,\ldots)$ labels a conjugacy class $C({\vec k})$ of $S_k$  since  ${\vec k}$ corresponds to a partition of $k$:
\ba
k=\sum_j jk_j.
\ea
The integers $z_{\vec k}\equiv \prod_j k_j!j^{k_j}$ encode the number of permutations $N(C({\vec k}))$ in the conjugacy class $C(\vec{k})$, which is given by $N(C({\vec k}))={k!/ z_{\vec k}}$. Also:
\ba
\hbox{Tr}_{\vec k}V\equiv \prod_j(\hbox{Tr}V^j)^{k_j}.
\ea
The function $f_{\vec k}(q,e^{-\vec{Q}\cdot \vec{t}})$ in (\ref{openGV}) can   be written in terms of the open GV invariants ${\hat N}_{Rg{\vec Q}}\in \Z$  \cite{Ooguri:1999bv,Labastida:2000yw}:
\ba
f_{\vec k}(q^n,e^{-n\vec{Q}\cdot \vec{t})}=\sum_{g=0}^\infty \ [n]^{2g-2} \prod_{j=1}^\infty[nj]^{k_j} 
 \sum_{\vec{Q}} \sum_{R }\chi_{R}(C(\vec k))
 {\hat N}_{Rg{\vec Q}}e^{-n\vec{Q}\cdot \vec{t}}.
 \label{openGVa}
\ea
As before $[a]\equiv q^{a/2}-q^{-a/2}$, $R$ is a representation of $S_k$ and\footnote{We recall that the representations of $U(P)$ and $S_k$ are both labeled by a Young tableau.}   of $U(P)$   labeled by a Young tableau $R$ and $\chi_{R}(C(\vec k))$ is the character  in the representation $R$ of $S_k$ for the conjugacy class $C({\vec k})$.
 Roughly speaking, the integers ${\hat N}_{Rg{\vec Q}}$  count\footnote{$g$ encodes the
quantum number under   $SO(2)$,   the rotation group in the  two non-compact directions.} the number of BPS multiplets wrapping the class\footnote{$H_2(X,L)$ denotes the relative homology group.}
${\vec \Sigma}\cdot {\vec Q}\in H_2(X,L,\Z)$ transforming in a representation $R$ of $U(P)$
and ending on the $D4$-branes wrapping $L$.

   Knowledge of the open GV invariants ${\hat N}_{Rg{\vec Q}}$ and the holonomy matrix $V$  corresponding to
   a $D$-brane configuration in $X$ determines using (\ref{openGV}) the open topological string partition function in $X$ to all orders  in perturbation theory.




\section{\texorpdfstring{Open strings in $X\ =\ $ closed strings in $X_b$}{Open strings in X = closed strings in Xb}}
 \label{open2closedGV-section}

We are now going to evaluate the open string partition function  in a Calabi-Yau $X$ (\ref{openGV})
  and show  that the resulting open+closed partition function in $X$   takes precisely the form of a closed string partition function (\ref{closedGV}) on a new Calabi-Yau manifold $X_b$! The physical interpretation of $X_b$ is as the Calabi-Yau space obtained by letting the $D$-branes in $X$ undergo a geometric transition. From the identification of partition functions we can compute the closed GV invariants\footnote{We note that ${\vec {Q}_b}\in \Z^{b_2(X_b)}$
while ${\vec Q}\in \Z^{b_2(X)}$. We shall see that if $R$ is parametrized as in  Figure \ref{table}, then $H_2(X_b,\Z)\simeq H_2(X,\Z)\oplus \Z^{2m}$.} $n_g^{\vec {Q}_b}(X_b)$ in  $X_b$ in terms of the open ${\hat N}_{Rg{\vec Q}}(X)$ and closed $n_g^{\vec Q}(X)$
GV invariants in $X$.


The open+closed topological string partition function in $X$ has a contribution from the open string sector living on the $D$-brane configuration under study and one from the closed string sector. Therefore, the partition function factorizes into two pieces
\ba
Z_{o+c}(X)=Z_{o}(g_s,t,V)\cdot Z_{c}(g_s,t),
\label{factor}
\ea
the first arising from world-sheets with boundaries while the second one from world-sheets without boundaries.  $n_g^{\vec Q}(X)$ determines $ Z_{c}(g_s,t)$ while ${\hat N}_{Rg{\vec Q}}(X)$ together with the holonomy of the gauge field determines $Z_{o}(g_s,t,V)$.
Since our goal is to show that the open+closed partition function  in $X$ (\ref{factor}) takes the form of a closed string partition function $Z_c(X_b)$, the main task is to show that
the open string contribution to (\ref{factor}) can be rewritten as a closed string amplitude. Of course, the
detailed form of the closed string partition function in $X_b$ will depend on the closed string partition function in $X$.

The open string partition function on such a $D$-brane configuration in $X$
is completely characterized by  the corresponding  open GV invariants in $X$ {\it and}
by specifying the holonomy of the gauge field ${\cal A}$ living on the $D$-brane configuration.
Since the $D$-branes wrap a Lagrangian submanifold
$L$ with $b_1(L)\neq 0$,
the $D$-brane amplitude depends on the gauge invariant\footnote{This is gauge invariance under closed string field theory gauge transformations, which act by \newline
${\cal J}\rightarrow {\cal J}+ d\Lambda,\
{\cal A}\rightarrow {\cal A}-\Lambda$.}
holonomy matrix
\ba
V=P\exp \left[-\left( \oint_\beta {\cal A}+\int_D {\cal J}\right)\right], \label{holo-mat}
\ea
where ${\cal J}$ is the complexified \Kahler form, $\beta\in H_1(L)$ and $D$ is a two-chain with $\partial D=\beta$.
  Geometrically,  the holonomy of the gauge field (\ref{holo-mat})  is gauge equivalent  to the ``position"\footnote{The position is defined only up to Hamiltonian deformations, which are gauge symmetries of the A-model open string field theory.}  of the branes  in $X$.
  Therefore, the  holonomy is part of the data that the open string theory depends on.

Following  \cite{Gomis:2006mv}, we turn on discrete values of the holonomy matrix  (\ref{holo-mat})
determined by a Young tableau $R$.
For  a configuration of  $P$ $D$-branes the holonomy matrix can be diagonalized
\ba
V\equiv U_R=\hbox{diag}\left(e^{-a_1},e^{-a_2},\ldots, e^{-a_P}\right),
 \label{holoa}
\ea
where the  eigenvalue $a_i$ corresponds to  the ``position'' of the $i$-th brane,  which  is given by \cite{Gomis:2006mv}
\ba
 a_i\equiv\oint_\beta {\cal A}_i +\int_D{\cal J}=g_s\left( R_i-i+P+\half\right), ~~ i=1,\ldots,P.
\label{periodsa}
\ea
 $R_i$ is the number of boxes in the $i$-th row of the Young tableau $R$:
 \begin{figure}[h]
 \begin{center}
\psfrag{R1}{$R_1$}
\psfrag{R2}{$R_2$}
\psfrag{RP-1}{$R_{P-1}$}
\psfrag{RP}{$R_P$}
\includegraphics[width=50mm]{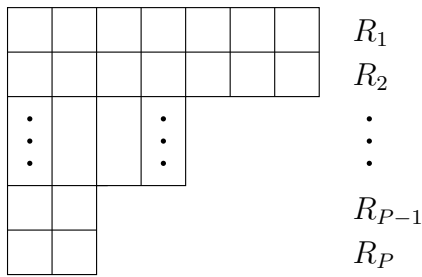}
\caption{A Young tableau $R$.  $R_i$ is the number of boxes in the $i$-th row. 
It satisfies $R_i\geq R_{i+1}$.}
\end{center}
\end{figure}


The explicit formula for the closed GV invariants in $X_b$ depends
on the closed GV invariants in $X$, the open GV invariants of the $D$-brane configuration in $X$ and on the holonomy
of the gauge field (\ref{holoa}) on the branes, which is determined by a Young tableau $R$.
The most interesting contribution to the formula  we derive for the closed GV invariants in   $X_b$ arises from the open string partition function of the brane configuration in $X$,
since $Z_{c}(g_s,t)$ in (\ref{factor}) already takes the form of a closed string partition function.

We start by performing our computations for the case when $X$ is the resolved conifold geometry. Apart from already capturing the closed string, bubbling Calabi-Yau interpretation of $D$-branes  in a simple setting, it also has interesting applications to knot invariants.
We find that the closed topological string partition function on certain bubbling Calabi-Yau manifolds are invariants of knots in $S^3$.



We   want to compute the open+closed topological string partition function on the resolved conifold geometry.
In order to define the open string partition function we must first specify a $D$-brane configuration in the resolved conifold giving rise to BRST-invariant boundary conditions on the string world-sheet, corresponding to branes wrapping a Lagrangian submanifold.
 The resolved conifold is an asymptotically conical  Calabi-Yau with base $S^2\times S^3$ and
 topology $R^4\times S^2$. One can construct a Lagrangian submanifold $L$ for every knot $\alpha$
 in the $S^3$ at asymptotic infinity \cite{Taubes:2001wk,koshkin}.
 We can then study the open string theory defined by $D$-branes wrapping these Lagrangian submanifolds, which have  topology
$L\simeq {\rm R}^2\times S^1$ and end on a knot $\alpha$ at asymptotic infinity.

We consider the open+closed string partition when $P$ $D$-branes wrap a Lagrangian submanifold $L$ associated to an arbitrary knot
$\alpha\subset S^3$. There are several contributions, from both the open and closed string sector.

The closed string contribution is well known \cite{Gopakumar:1998ii,faber}:
\ba
Z_{c}(g_s,t)=M(q)\cdot \exp\left(-\sum_{n=1}^\infty \oo{ n\: [n]^{2}} e^{-nt}\right).
\label{closedcon}
\ea
Comparing with the general formula for the closed string partition function in terms of
the closed GV invariants (\ref{closedGV}) one finds that there is a unique non-vanishing closed GV invariant in the resolved conifold geometry, given by $n_0^1=-1$. For the resolved conifold  geometry $b_2(X)=1$ -- and $\chi(X)=2$  -- and $t=\int_{S^2} {\cal J}$ parametrizes the complexified size of the $S^2$.

The open string contribution to the partition function has several pieces. One contribution is  captured by the open string partition function in
(\ref{openGV}).  The holonomy of the gauge field (\ref{holo-mat}) around the non-contractible one-cycle
$\beta$ in the Lagrangian $L$,
--  labeled by the knot $\alpha$\footnote{Note that the knot $\alpha\subset S^3 $ is contractible in $L$.} -- must be given to completely specify the $D$-brane configuration, and the corresponding open string theory. This is because the holonomy of the gauge field determines the positions of the $D$-branes up to Hamiltonian deformations\footnote{
A Hamiltonian deformation is generated by a vector $v$ in the normal bundle of $L$ of the form $v^\mu=(w^{-1})^{\mu\nu}\partial_\nu f$ for arbitrary $f$, where $w_{\mu\nu}$ is the \Kahler form of the symplectic manifold $X$.} \cite{Gomis:2006mv}, which are gauge symmetries of the A-model open string field theory.
Following \cite{Gomis:2006mv} we now turn on a non-trivial holonomy $V=U_R$ (\ref{holo-mat})
labeled by a Young tableau $R$ (\ref{holoa}, \ref{periodsa}). Turning on a non-trivial holonomy has the effect of separating the branes, and therefore making the off-diagonal open strings massive. Integrating these fields out also contributes to the open string  amplitude on the $D$-brane configuration. Combining the various terms we have that the complete open string partition function is given by
\ba
Z_{o}(g_s,t,V=U_R)=\exp\hskip-3pt\left(\sum_{n=1}^\infty\oo{n}\hskip-3pt\left[ -\hskip-3pt\sum_{1\leq i<j\leq P} e^{-n(a_i-a_j)}+\sum_{\vec k}\oo{z_{\vec k}}f_{\vec k}(q^n,e^{-nt}) \Tr_{\vec
k} U_R^n
\right]\hskip-3pt\right)
\label{openampl}
\ea
where
\ba
\exp\left(-\sum_{n=1}^\infty \oo{ n}\sum_{1\leq i<j\leq P} e^{-n(a_i-a_j)}\right)=\prod_{1\leq i<j\leq P}(1-e^{-(a_i-a_j)})
\label{vander}
\ea
 arises by integrating out the off-diagonal massive open strings.
 From a world-sheet perspective this last contribution arises from world-sheet annuli
 connecting the various $D$-branes\footnote{
Though (\ref{vander}) looks like a fermion determinant if we naively apply the argument of \cite{Ooguri:1999bv},
the massive open string is a boson.
The argument does not really apply because the open string is not localized along an $S^1$.
It instead applies to the related toric situation where an open string stretches
between one brane $D_1$ wrapping $L_1$ and another $D_2$ wrapping $L_2$.
Here $L_1$ and $L_2$ are two Lagrangians that can combine and move off to infinity \cite{Aganagic:2000gs}.
The open string is localized along $L_1\cap L_2=S^1$, and the argument of \cite{Ooguri:1999bv}
implies that it contributes the bosonic determinant $1/(1-e^{-\Delta a})$.
If $D_1$ and $D_2$ both wrap $L_1$ (or $L_2$),
the contribution from the stretched open string is the inverse $(1-e^{-\Delta a})$,
which appears in (\ref{vander}).
We thank M. Aganagic for explaining this to us.}.

%
%

By combining the closed string partition function
(\ref{closedcon}) with the open string partition function (\ref{openampl}), 
  we find that the open+closed partition function for a configuration of $P$ $D$-branes  wrapping a Lagrangian submanifold $L$ in the resolved conifold is given by:
\ba
Z_{o+c}=M(q)\exp\hskip-3pt\left(\sum_{n=1}^\infty\oo{n}\hskip-3pt\left[ -\f{e^{-nt}}{ [n]^2}-\hskip-10pt\sum_{1\leq i<j\leq P} e^{-n(a_i-a_j)}+\hskip-2pt\sum_{\vec k}\oo{z_{\vec k}}f_{\vec k}(q^n,e^{-nt}) \Tr_{\vec
k} U_R^n
\right]\hskip-3pt\right).
\label{open+closed}
\ea

 The first step in identifying the open+closed string partition function in (\ref{open+closed}) as a purely closed string amplitude is to write the contribution from the off-diagonal massive open strings in
 (\ref{vander})  as a closed string world-sheet instanton expansion. For this purpose,  it is convenient to parametrize the Young tableau using the coordinates in Figure \ref{table}. Then the following useful identity can be derived (see Appendix \ref{rewriteform})
 \ba
&&\xi(q)^{P}\exp\left(-\sum_{n=1}^\infty\oo n\sum_{1\leq i<j\leq P} e^{-n(a_i-a_j)}\right) \nn\\
&=&\hskip-4pt\ M(q)^m\exp\left( \sum_{n=1}^\infty \oo{n\: [n]^{2}} \left[\sum_{1\leq I\leq J\leq 2m-1}  (-1)^{J-I+1}
e^{-n(t_I+t_{I+1}+...+t_J)}\right]\hskip-2pt\right),
\label{cyl-amp}
\ea
where we have identified
\ba
t_I=g_sl_I\qquad     I=1,\ldots,2m
\label{sizes}
\ea
with $l_I$ being the coordinates of the Young tableau in Figure \ref{table}.
$M(q)$ is the MacMahon function (\ref{mac}) and  $\xi(q)=\prod_{j=1}^\infty(1-q^j)^\mo$. In this way  we have written the contribution from open string world-sheets with annulus topology as a closed string instanton expansion.

We can also derive the following formula for the holonomy of the gauge field on the branes (see Appendix \ref{rewriteform})
\ba
\Tr_{\vec k} U_R^n=
\prod_{j=1}^\infty\left(\f{\sum_{I=1}^{m}   e^{-njT_{2I-1}}- e^{-njT_{2I}} }{ [nj]}\right)^{k_j},
\label{tracerep}
\ea
 with $U_R$ given in (\ref{holoa}, \ref{periodsa}).
 Here
 \ba
 T_{I}=\sum_{J=I}^{2m}t_J
 \label{bigts}
 \ea
 and $[nj]=q^{nj/2}-q^{-nj/2}$, where $q=e^{-g_s}$. Therefore, the contribution of  the holonomy matrix  to the open string amplitude
 (\ref{openampl})  also takes the form of a world-sheet instanton expansion with
 \Kahler parameters $t_I$, with $I=1,\ldots,2m$.
 For later purposes it is convenient to introduce the notation
 \ba
 e^{-nT_o}\equiv
 (e^{-nT_1},e^{-nT_3},...,e^{-nT_{2m-1}}), ~~   e^{-nT_e}\equiv
 (e^{-nT_2},e^{-nT_4},...,e^{-nT_{2m}}).
 \ea

A crucial step in uncovering the closed string  interpretation of open string amplitudes in
topological string theory is to use the following identity
(proven in Appendix  \ref{integralityapp} using CFT techniques,
which are reviewed in Appendix \ref{rewriteoperator})
\ba
\sum_{\vec k}\oo{z_{\vec k}}\chi_{R_1}(C(\vec k))
\prod_{j=1}^\infty \left(
\sum_{I=1}^{m}
\lambda_I{}^j-\sum_{I=1}^m \eta_I{}^j\right)^{k_j}\hspace{-3mm}=\hspace{-3mm}
\sum_{R_1,R_2,R_3}(-1)^{|R_3|}N^{R_1}_{R_2R_3} s_{R_2}(\lambda) s_{R_3^T}(\eta), \label{identity}
\ea
where $\lambda=(\lambda_I)$ and $\eta=(\eta_I)$ with $I=1,...,m$ are arbitrary variables.
The left hand side of (\ref{identity}) enters in the parametrization of  the open string partition function  in (\ref{openGV})
 by using (\ref{openGVa}).
The  symbol $N^{R_1}_{R_2R_3}$ denotes the Littlewood-Richardson coefficients of $U(P)$,
which determine the number of times the representation $R_1$  of $U(P)$ appears
in the tensor product of  representations $R_2$ and $R_3$ of $U(P)$.
$R_3^T$ is the representation of $U(P)$ obtained by transposing the Young tableau $R_3$.
Finally, $s_{R}(x)$ is a  Schur polynomial of $U(m)$, which is labeled  by a Young tableau $R$.
It is defined by taking the trace\footnote{In terms of the fundamental representation, we have that
$\hbox{Tr}_R X=\sum_{{\vec k}}\oo{ z_{\vec k}}\chi_R(C({\vec k})) \prod_j (\hbox{Tr} X^j)^{k_j}$.} in the representation $R$
\ba
s_R(x)\equiv \hbox{Tr}_R X,
\ea
 where $X$ is an
 $m\times m$  diagonal matrix 
 with entries $X\equiv{\rm diag} (x_1,\ldots,x_m)$.
%

 We can now use  (
 \ref{tracerep},
 \ref{identity})\footnote{For the resolved conifold there is only one K\"ahler modulus, which  we denote by $t$.} to write the second  term in the open string partition function
 on the resolved conifold  (\ref{openampl}) as follows:
\ba
&&\sum_{\vec k}\oo{z_{\vec k}}f_{\vec k}(q^n,e^{-nt}) \Tr_{\vec
k} U_R^n \nn\\
&=&\sum_{g=0}^\infty \sum_{Q\in \Z}\sum_{R_1,R_2, R_3}\f{1}{ [n]^{2-2g}} {\hat N}_{R_1gQ}
(-1)^{|R_3|} {N}^{R_1}_{R_2R_3} s_{R_2}(e^{-nT_o})s_{R_3^T}(e^{-nT_e})
e^{-nQt}.\hspace{5mm}
\label{integrality}
\ea
 We note that the factor $[nj]^{k_j}$  in the definition of $f_{\vec k}$ in (\ref{openGVa}) precisely cancels
 with an identical factor in (\ref{identity}).

Therefore, we have proven that the open+closed partition function on the resolved conifold (\ref{open+closed})  can be written as follows\footnote{
In writing this, we have dropped an ambiguous factor proportional to $\xi(q)$,
which does not affect the answer to any order in perturbation theory \cite{Saulina:2004da}.}:
 \ba
Z_{o+c}&=&
M(q)^{m+1}\exp\Bigg(\sum_{g=0}^\infty\sum_{n=1}^\infty  \f{1}{ n\ [n]^{2-2g}}\Bigg[
-\delta_{g0} e^{-nt}
\nn\\
&&~~~+\delta_{g0} \sum_{1\leq I\leq J\leq 2m-1} (-1)^{J-I+1}
e^{-n(t_I+t_{I+1}+...+t_J)} 
\Bigg.
\nn\\
&&~~~+\sum_{Q\in \Z}
\sum_{R_1,R_2, R_3} {\hat N}_{R_1gQ}
(-1)^{|R_3|} N^{R_1}_{R_2R_3} s_{R_2}(e^{-nT_o})s_{R_3^T}(e^{-nT_e})
e^{- nQt}\Bigg]\Bigg).\hspace{10mm}
\label{open+closeda}
 \ea

 A quick glance at the formula for the closed topological string partition function in terms of
 closed GV invariants (\ref{closedGV}) confirms that the open+closed partition function
 in the resolved conifold (\ref{open+closeda}) takes precisely the form of a closed string partition function on a different Calabi-Yau space $X_b$. Moreover, by using  that the Littlewood-Richardson coefficients $N^{R_1}_{R_2R_3}$ are integers and that a Schur polynomial $s_R(M)$ is a symmetric polynomial of the eigenvalues of $M$
 with integer coefficients,
 we can  conclude  that the coefficients in (\ref{open+closeda}) have  the correct integrality properties  for  a closed string amplitude parametrized by closed GV invariants. Therefore, we have proven that the open+closed string partition function on the resolved conifold
takes precisely the form  of a closed string partition function in
another Calabi-Yau $X_b$  with  the correct {\it integrality} properties!

It follows from the expression in (\ref{open+closeda}) that the Calabi-Yau manifold $X_b$ has different topology than the Calabi-Yau space we started with.
In fact, by looking at the exponent of $M(q)$ in (\ref{open+closeda}) we have shown that $\chi(X_b)=2m+2$. The appearance of the extra cycles has a simple physical intepretation. The branes in the resolved conifold have undergone a geometric transition and have been replaced by flux. Fluxes in the topological string correspond precisely to non-trivial periods of the complexified \Kahler form ${\cal J}$. In this picture, the original branes disappear and leave behind a collection of non-contractible cycles on which their flux is supported.
It is this picture that warrants the description of $X_b$ as a bubbling Calabi-Yau.

It is now straightforward to extend the computation of the open+closed partition function to an arbitrary Calabi-Yau $X$. The open+closed partition function of a $D$-brane configuration in $X$ is given by:
\ba
Z_{o+c}&=&
M(q)^{\f{\chi(X)+2m}{2}}\exp\Bigg(\sum_{g=0}^\infty\sum_{n=1}^\infty  \f{1}{ n\ [n]^{2-2g}}\Bigg[ \sum_{\vec{Q}} n_g^{\vec{Q}} e^{-n\vec{Q}\cdot \vec{t}}
\nn\\
&&~~
+\delta_{g0}\sum_{1\leq I\leq J\leq 2m-1} (-1)^{J-I+1}
e^{-n(t_I+t_{I+1}+...+t_J)} \nn\\
&&~~
+\sum_{{\vec Q}}
\sum_{R_1,R_2, R_3}{\hat N}_{R_1g{\vec Q}}
(-1)^{|R_3|} N^{R_1}_{R_2R_3} s_{R_2}(e^{-nT_o})s_{R_3^T}(e^{-nT_e})
e^{-n{\vec Q}\cdot{\vec t}}\Bigg]\Bigg).\hspace{20mm}
\label{open+closedb}
 \ea
 The integers $n_g^{\vec{Q}}$ are the  closed GV invariants in $X$,
 which determine the closed string partition function in $X$,
 where now ${\vec Q}\in \Z^{b_2(X)}$. As before, the integers ${\hat N}_{R'g{\vec Q}}$ are the open GV invariants of the $D$-brane configuration in $X$. Just as in the case when $X$ is the resolved conifold,
 the open+closed partition function (\ref{open+closedb}) takes  precisely
 the form of a closed string partition function in $X_b$ (\ref{closedGV}), with integral closed GV invariants. This explicitly shows that the physics of $D$-branes in $X$ can be either described by open+closed string theory in $X$ or equivalently by closed string theory on a topologically different manifold $X_b$. Showing that the open+closed string theory in $X$ has a closed string interpretation in $X_b$ does not rely on explicitly knowing the open and closed GV invariants in $X$. Nevertheless, since the open and closed  partition function take a very particular form in topological string theory -- being parametrized by integer invariants --, we can show that we the  open string amplitude in $X$  takes the form of a closed string  amplitude in $X_b$.

%

We can explicitly compute the closed GV invariants $n_g^{\vec Q_b}(X_b)$ in $X_b$ in terms of the open ${\hat N}_{Rg{\vec Q}}$ and closed $n_g^{\vec Q}$ GV invariants in $X$
by comparing the open+closed string partition function in $X$ (\ref{open+closedb}) with the general expression for the
closed string partition function in topological string theory (\ref{closedGV}). By matching the two series we get:
\ba
&&\sum_{\vec Q_b}n_g^{\vec Q_b}(X_b) e^{-\vec Q_b\cdot\vec t} \nn\\
&&=\sum_{\vec Q}n_g^{\vec Q} e^{-\vec Q \cdot\vec t}+\delta_{g0}
\sum_{1\leq I\leq J\leq 2m-1}  (-1)^{J-I+1}
e^{-t_I-t_{I+1}-...-t_J}\nn\\
&&+ \sum_{{\vec Q}}\sum_{R_1R_2R_3}
 {\hat N}_{R_1g{\vec Q}} e^{-{\vec Q}\cdot\vec t}(-1)^{|R_3|}N^{R_1}_{R_2R_3}
s_{R_2}(e^{-T_o}) s_{R_3^T}(e^{-T_e}).
 \label{open2closedGV}
\ea
 By  comparing the two series one can explicitly calculate $n_g^{\vec Q_b}(X_b)$ in terms of ${\hat N}_{R_1g{\vec Q}}$ and $n_g^{\vec Q}$. In Appendix  \ref{closed2open-appendix},
 we rewrite (\ref{open2closedGV})
in  a form in which it is easy to obtain the closed  GV invariants in $X_b$ from the open and closed GV invariants in $X$.

\eject

%
\medskip
\medskip
\noindent{\it Continuous v.s. discrete holonomies and framing dependence}
\medskip
\medskip

Holonomy taking discrete values plays a crucial role in the discussion
in \cite{Gomis:2006mv} and this paper.
On the other hand, most topological string literature starting with \cite{Ooguri:1999bv}
has assumed that holonomy takes continuous values.
It is natural to ask what is the relation between the two pictures.

Our proposal is that the partition function in one picture with one framing is
a linear combination of partition functions in the other picture with an appropriate framing.
We now explain this statement in some detail.
Let us assume that the Lagrangian submanifold $L$ the $D$-branes wrap has topology
of ${\rm R}^2\times S^1$, which can be regarded as solid torus.
At asymptotic infinity, the geometry is a cone over $T^2$.
Given $L$, there is a unique one-cycle of $T^2$ that is contractible in $L$.
In fact, as one moves from one point to another one in the quantum moduli space
of such $D$-branes,  the original contractible cycle can become non-contractible
while another cycle becomes contractible.
In other words, the quantum moduli space
contains topologically distinct Lagrangian submanifolds that are related by a flop.
The open string partition function $Z_o(g_s,V;f_1)$ is a wave function
in Chern-Simons theory on the $T^2$ at infinity.
The definition of the wave function involves framing($=$the choice of polarization) $f_1$,
i.e., the choice of variables corresponding to a coordinate and its conjugate momentum.
In the case of Chern-Simons theory on $T^2$, polarization is fixed
by choosing a pair of symplectic generators $(\alpha,\beta)$ such that $\#(\alpha\cap \beta)=1$.
$\oint_\alpha \Acal$ plays the role of a coordinate and $\oint_\beta \Acal$ the role of the conjugate momentum.
$g_s$ plays the role of the Planck constant \cite{Elitzur:1989nr}.
The conventional picture of holonomy is such that $V\sim\exp-\oint_\alpha \Acal$,
where $\alpha$ is a non-contractible cycle.
Since $\oint_\alpha \Acal$ is a periodic variable, the conjugate momentum $\oint_\beta\Acal$
gets quantized in units of $g_s$.
A basis state $|R\rangle$ of the Hilbert space in our polarization is labeled by a Young tableau $R$,
and this state corresponds to $\exp-\oint_\beta \Acal=U_R$ \cite{Elitzur:1989nr}.
On the other hand, the state in which $\exp -\oint_\alpha\Acal$ equals $V$ is $|V\rangle=\sum_R \Tr_R V |R\rangle$.
We expect that there is a point in the moduli space where $\alpha$ is a non-contractible cycle of $L$.
We also expect that the two open string partition functions are related as
$Z_o(g_s, V;f_1)=\sum_R \Tr_R V Z_o(g_s,U_R;f_2)$ with appropriate framing $f_2$.
This is indeed what happens for the $D$-branes corresponding to unknot in $S^3$
up to normalization and a shift in the \Kahler modulus \cite{Okuda:2007ai}.
%

\medskip
\medskip
\noindent{\it Knot invariants from closed strings in bubbling Calabi-Yau manifolds}
\medskip
\medskip

In \cite{Gomis:2006mv} we identified the $D$-brane configurations\footnote{
The convention for the distinction of brane/anti-brane here is the opposite of \cite{Gomis:2006mv}.
} in the resolved conifold $X$ corresponding 
to a Wilson loop in $U(N)$ Chern-Simons theory on $S^3$. 
The brane configuration depends
on the  knot $\alpha\subset S^3$ and on the choice of a representation $R$ of $U(N)$,  which is the data on which  the Wilson loop depends on (see  \cite{Gomis:2006mv} for the details of the brane  configuration).

\begin{figure}[ht]
\centering
\begin{tabular}{cc}
\psfrag{t}{$t$}
\includegraphics[scale=.6]{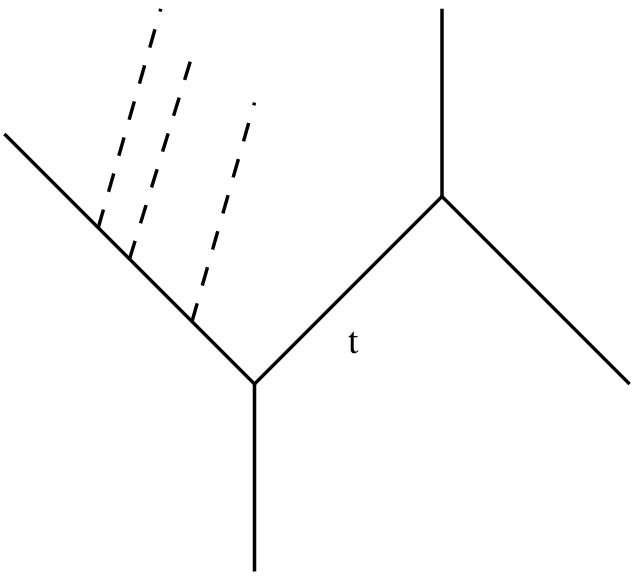}
&
\psfrag{t}{$t$}
\psfrag{t1}{$t_1$}
\psfrag{t2}{$t_2$}
\psfrag{t2m-2}{$t_{2m-2}$}
\psfrag{t2m-1}{$t_{2m-1}$}
\psfrag{t2m}{$t_{2m}$}
\includegraphics[scale=.6]{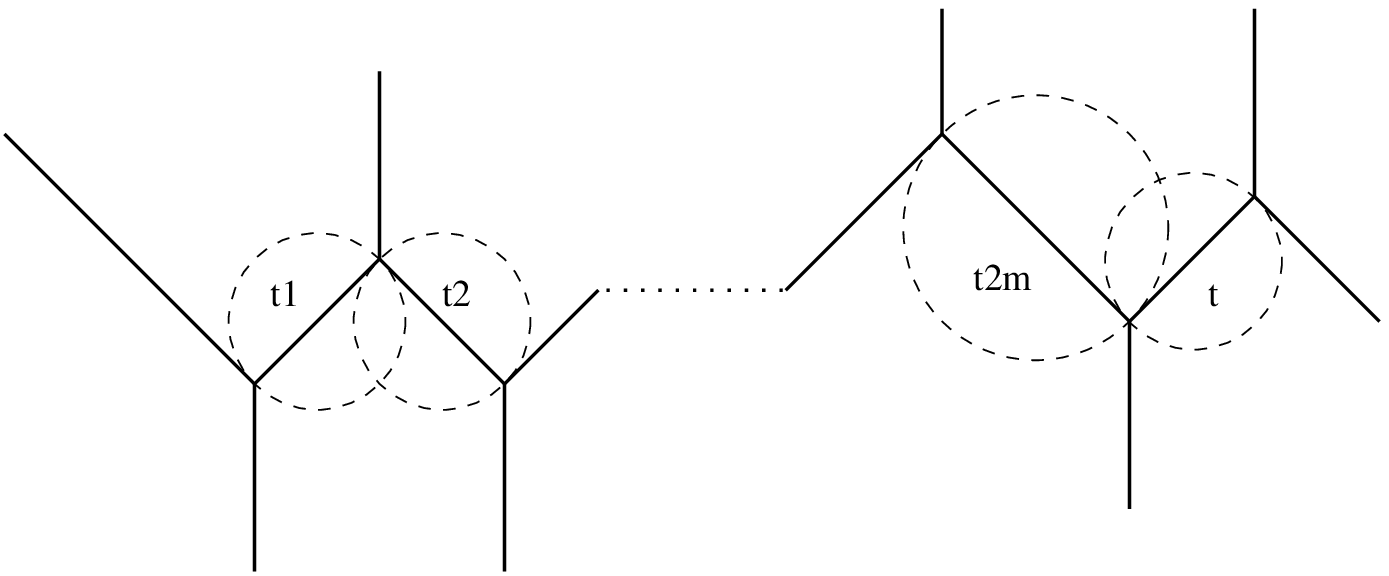}
\\
(a)&(b)
\end{tabular}
\caption{
(a) The resolved conifold and $D$-branes with holonomy $U_R$  inserted on an outer edge.
(b) The bubbling Calabi-Yau $X_b$ after geometric transition of the $D$-branes.
The \Kahler moduli are given by $t_I=g_s l_I$, $I=1,...,2m$, where $l_I$ are defined in Figure \ref{table}.
}
\label{knot-brane-bubbling}
\end{figure}

This identification was explicitly verified for the case when $\alpha$ is the unknot
and for an arbitrary representation $R$.
In addition,
we noticed that the $D$-brane configuration\footnote{As explained in  \cite{Gomis:2006mv}, a given Wilson loop can be represented either in terms of $D$-branes or anti-branes in the resolved conifold, in an analogous fashion to the AdS description of half-BPS Wilson loops
\cite{Gomis:2006sb}. Both brane configurations give rise to the same bubbling Calabi-Yau $X_b$.}
 in the resolved conifold corresponding to
the unknot and for arbitrary representation $R$, shown in Figure \ref{knot-brane-bubbling}(a), could be given a purely closed string interpretation in terms of the closed string partition function
on a bubbling Calabi-Yau $X_b$ of ladder type, shown in Figure \ref{knot-brane-bubbling}(b).
More concretely, we showed  that \cite{Gomis:2006mv}
\ba
\left\langle \hbox{Tr}_R P \exp-\oint_\alpha A\right\rangle=Z_{o+c}(X)=Z_c(X_b),
\ea
where
\ba
Z_{o+c}(X)= M(q)
\exp\Bigg(\sum_{n=1}^\infty\oo n \Bigg[-\frac{e^{-n{t}}}{[n]^2} -\sum_{i<j}e^{-n(a_i-a_j)}+ \sum_{i=1}^P
 \frac{e^{-na_i}-e^{-n({t}+a_i)}}{[n]}\Bigg]\Bigg)
\label{conifold-brane-amplitude}
\ea
is the open+closed string partition function in the resolved conifold $X$, and
\ba
Z_c(X)&=&
M(q)^{m+1}\exp\sum_{n=1}^\infty \oo{n[n]^2}\left(
-\sum_{1\leq I\leq 2m+1} e^{-nt_I}
+\sum_{1\leq I\leq2m} e^{-n(t_I+t_{I+1})}\right.\nn\\
&&\left.~~
-\sum_{1\leq I\leq 2m-1} e^{-n(t_I+t_{I+1}+t_{I+2})}
...
- e^{-n(t_1+...+t_{2m+1})}
\right) \label{conifold-bubbling-amplitude}
\ea
 is the closed string partition function in $X_b$ with $t_{2m+1}\equiv t$.
The equality $Z_{o+c}(X)=Z_c(X_b)$ is of course the special case
of the result in the present paper.
By comparing (\ref{conifold-brane-amplitude}) with (\ref{open+closeda}),
we see that $\hat N_{\yng(1)\hspace{.7mm},g=0,Q=0}=1$ and $\hat N_{\yng(1)\hspace{.7mm},g=0,Q=1}=-1$ are the only
non-zero open GV invariants.
It can be seen that (\ref{conifold-bubbling-amplitude}) agrees with (\ref{open+closeda}).
One consequence of this identification is
that closed topological string theory  on bubbling Calabi-Yau
manifolds $X_b$  yield knot invariants for the unknot.

In this paper we have shown that {\it any} brane configuration in a Calabi-Yau manifold -- so in particular in the resolved conifold --  has a  purely closed string interpretation. Since we know \cite{Gomis:2006mv} which brane configuration corresponds to a  Wilson loop for arbitrary knot $\alpha$ and representation $R$, we can associate to the bubbling Calabi-Yau obtained from this brane configuration a knot.
This set of connections uncovers an interesting relation
between closed GV invariants in bubbling Calabi-Yau manifolds $X_b$ and invariants of knots in $S^3$.
It implies that the closed string partition function on appropriate bubbling Calabi-Yau manifolds $X_b$
are invariants of knots on $S^3$.



\section{Geometric transitions in toric Calabi-Yau's}\label{toricCY}

In this section we study the geometric transitions giving rise to bubbling Calabi-Yau
manifolds in the set-up of toric Calabi-Yau manifolds.
In addition to the general picture of geometric transitions presented in the previous
section, here we are able to concretely identify both the $D$-brane configurations and the bubbling
Calabi-Yau manifolds.
We explain how these geometric transitions can be understood by a combination
of complex structure deformation and a local version of conifold transition.
Furthermore we explicitly show, by using the topological vertex techniques,
that the open string partition function in a given $D$-brane configuration
is precisely the closed string partition function in the corresponding bubbling Calabi-Yau.


%

\subsection{
Local Gopakumar-Vafa duality}\label{localGV}

\begin{figure}[ht]
\centering
\begin{tabular}{cccc}
\psfrag{f}{$f$}
\includegraphics[scale=.25]{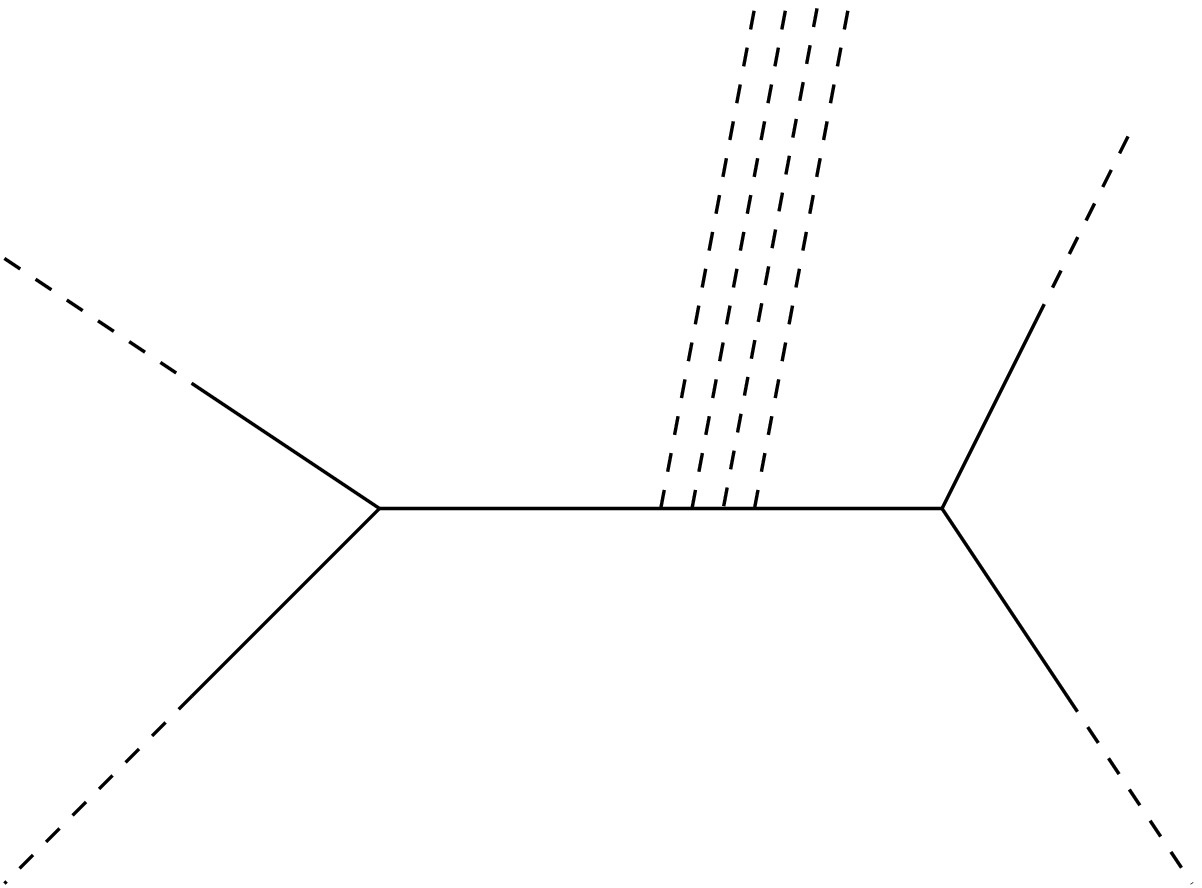}
&
\psfrag{S3}{$S^3$}
\includegraphics[scale=.25]{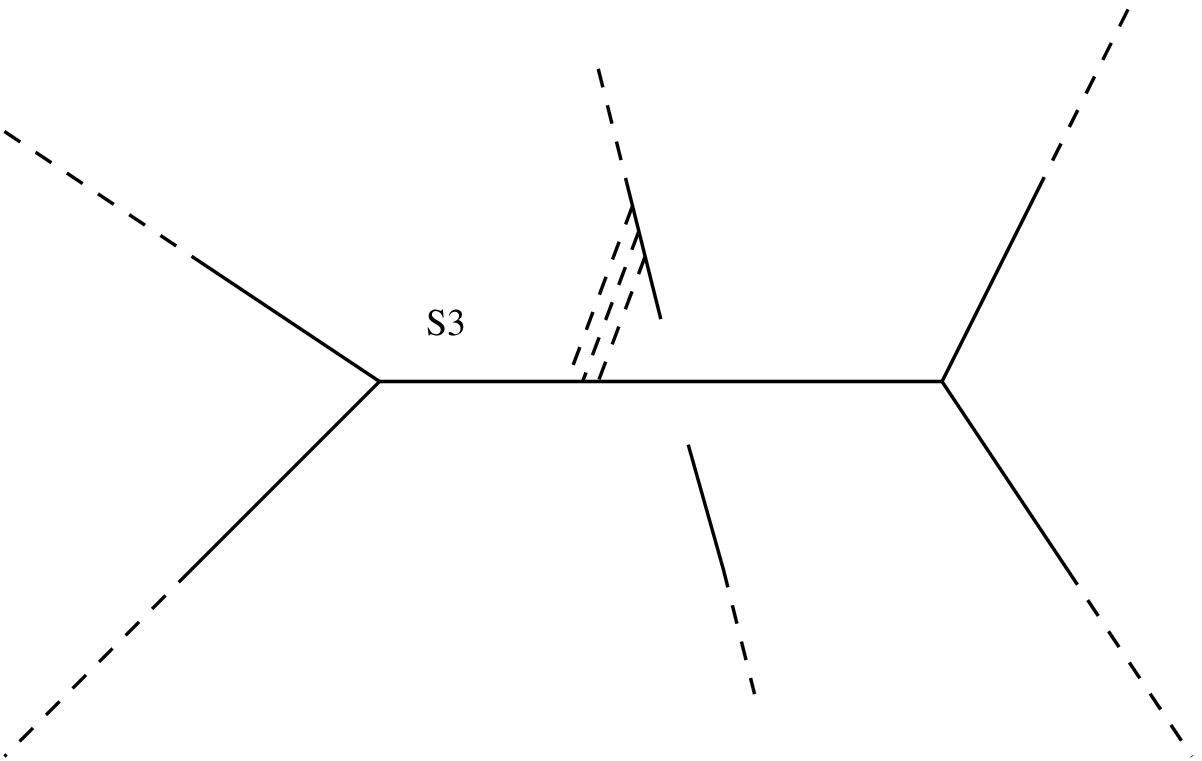}
&
\includegraphics[scale=.25]{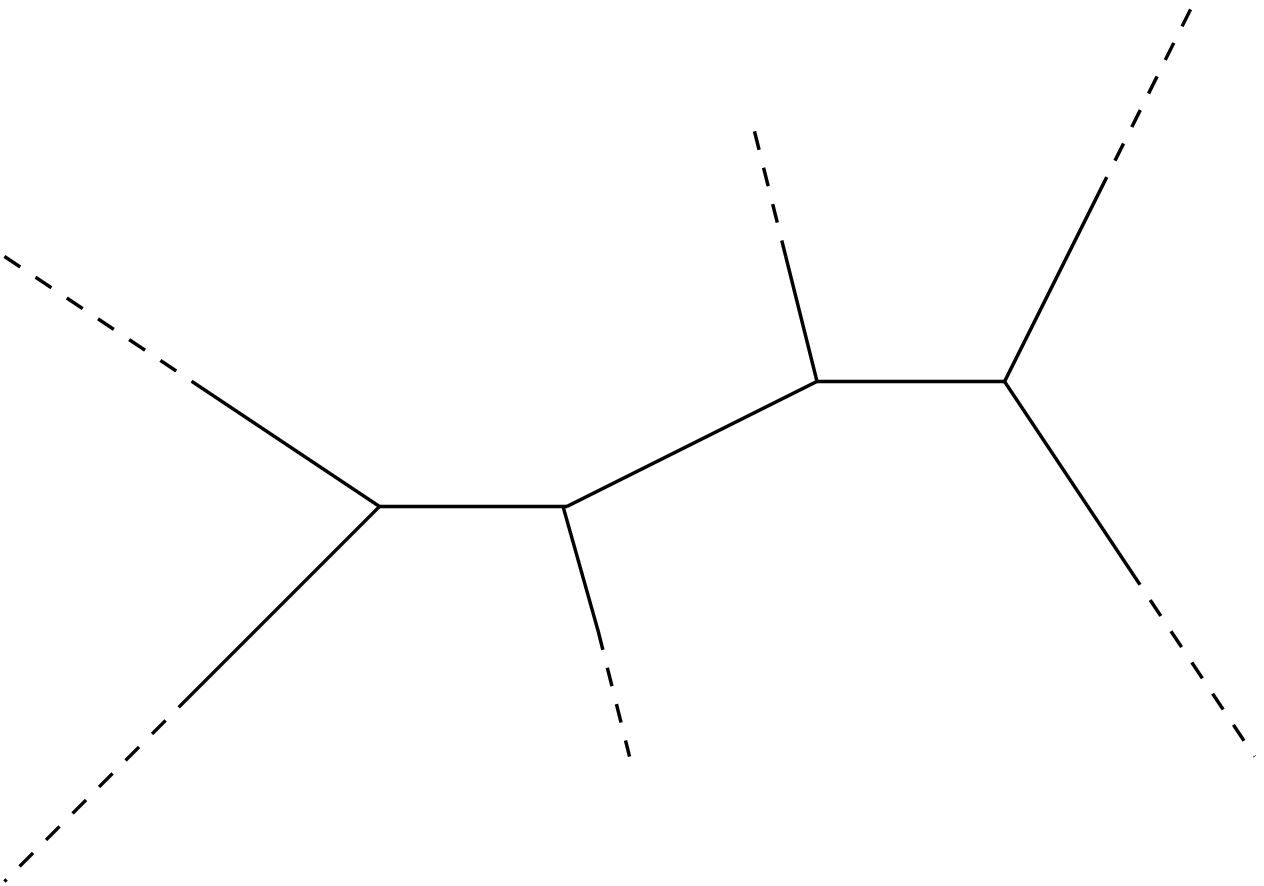}
&
\includegraphics[scale=.25]{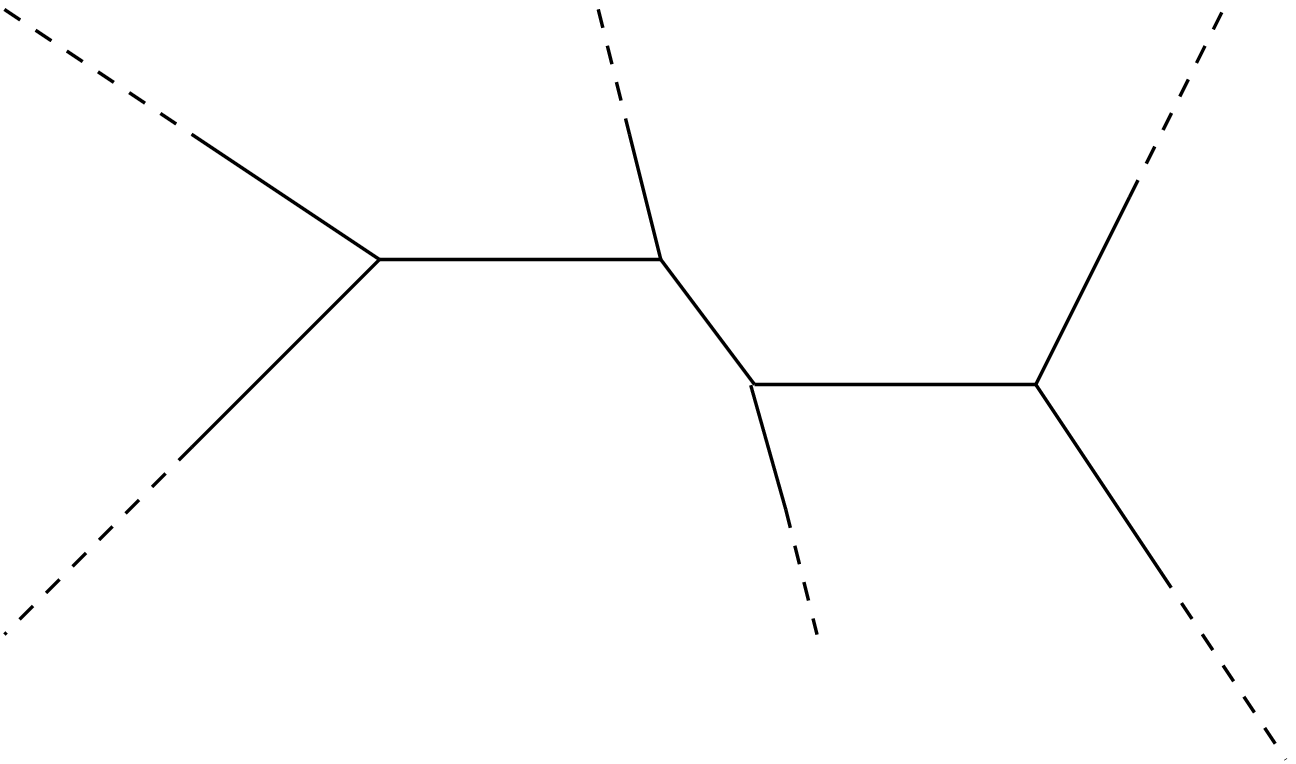}
\\
(a)&(b)&(c)&(d)
\end{tabular}
\caption{
(a) Non-compact $D$-branes (dashed lines ending on edges) in a toric Calabi-Yau manifold.
The framing of the branes is specified by a vector $f$.
(b) The geometry can be modified without changing the amplitude while
making the brane world-volume a compact $S^3$.
(c) The compact branes get replaced by a new 2-cycle upon geometric transition.
(d) Geometric transition of anti-branes produces a flopped geometry.
}
\label{compactification}
\end{figure}

Take an arbitrary toric Calabi-Yau manifold specified by a toric diagram.
Let us focus on one of the edges.
Without losing generality we assume that it is an internal edge\footnote{
By making the internal edge infinitely long one can trivially make it
external.}.
Consider $m$ non-compact branes wrapping a Lagrangian submanifold as shown in
Figure \ref{compactification}(a).
The submanifold has the topology of ${\rm R}^2\times S^1$, and
preserves an $U(1)^2\subset U(1)^3$ symmetry.
As explained in \cite{Aganagic:2003db}, it is possible to modify
the geometry so that the new geometry has a compact 3-cycle of $S^3$ topology
in  the edge\footnote{
In fact there is an infinite family of such modifications labeled by an integer $p$.
$p$ specifies framing of the non-compact branes as well as
the orientation of the new line in \ref{compactification}(b).
}.
Near the $S^3$ the local geometry is that of the deformed conifold.
The new geometry is not toric, but has the structure of an $\R\times T^2$ fibration \cite{Aganagic:2002qg}.
By a  complex structure deformation that makes the $S^3$ infinitely large,
one recovers the original toric Calabi-Yau.
The A-model amplitude is invariant under the complex structure deformation.

The $m$ branes now wrap the $S^3$
as shown in Figure \ref{compactification}(b).
In the limit of infinite $S^3$ size we get $m$ non-compact $D$-branes
ending on the edge in the original geometry, see Figure \ref{compactification}(a).
In the original geometry, the non-compact Lagrangian submanifold has
topology of ${\rm R}^2\times S^1$,
which we regard as a solid torus.
In particular it has a non-contractible $S^1$ cycle.
The non-compact Lagrangian is compactified to $S^3$ in the modified geometry.
If we focus on the Lagrangian alone, compactification is achieved by gluing another copy of the solid torus to the first copy
after applying the $S\in SL(2,\Z)$ transformation on the $T^2$ boundary.
The non-contractible $S^1$ becomes contractible in the new copy.
The Chern-Simons path integral on the new copy of the solid torus prepares a state on $T^2$,
which is the ground state because we insert no Wilson loop.
After the $S$ transformation, the ground state induces certain holonomy
along the $S^1$ proportional to the Weyl vector of $U(m)$ \cite{Elitzur:1989nr}:
\ba
-\oint {\cal A}= {\rm diag} \left(g_s\left[-i+\half+\f{m}{2}\right]\right)_{i=1}^m. \label{holonomy-ground-state}
\ea

We  now apply the local Gopakumar-Vafa duality \cite{Aganagic:2001ug}
to the branes wrapping the $S^3$.
The $m$ branes disappear and get replaced by a 2-cycle of topology $S^2$
with complexified \Kahler  modulus $g_s m$.
The local geometry is that of the resolved conifold with \Kahler parameter $g_sm$.
See Figure (c).
This makes clear that we need discrete values of holonomy on the branes to have
geometric transition\footnote{
Branes with continuous values of the holonomy on an edge are a superposition (integral transform) of
branes with discrete values of the holonomy ending on another edge \cite{Gomis:2006mv}.
The integral transform accounts for the change of polarization of Chern-Simons theory on $T^2$.
}.

If replace the branes by anti-branes we obtain a flopped geometry (Figure \ref{compactification}(d)).

\subsection{Geometric transition of branes in toric Calabi-Yau's}\label{geom-trans-toric}

We now verify our proposal for the  geometric transition described above.
This is done by showing, using the topological vertex formalism \cite{Aganagic:2003db},
that non-compact branes and anti-branes with certain discrete values of holonomy can be replaced
by  geometries.
As in much of recent literature we redefine $q\ra q^\mo$ relative to \cite{Aganagic:2003db}\footnote{
This is to ensure that infinite power series that appears in amplitudes involve positive powers of $q$.
Such convention is more natural in relation to the quantum foam picture \cite{Okounkov:2003sp,Iqbal:2003ds}.}.
Basic facts about the topological vertex are summarized in Appendix \ref{top-vert-app}.

\begin{figure}[t]
\centering
\begin{tabular}{ccc}
\psfrag{R}{$R$}
\psfrag{R1}{$R_1$}
\psfrag{R2}{$R_2$}
\psfrag{R3}{$R_3$}
\psfrag{R4}{$R_4$}
\psfrag{QL}{$Q_L$}
\psfrag{QR}{$Q_R$}
\psfrag{V}{$V$}
\psfrag{a1}{$a+\half g_s$}
\psfrag{a2}{
}
\psfrag{a3}{
}
\psfrag{v}{$v$}
\psfrag{v1}{$v_1$}
\psfrag{v2}{$v_2$}
\psfrag{v3}{$v_3$}
\psfrag{v4}{$v_4$}
\psfrag{v5}{$v_5$}
\psfrag{f}{$f$}
\psfrag{t}{$t$}
\includegraphics[scale=.55]{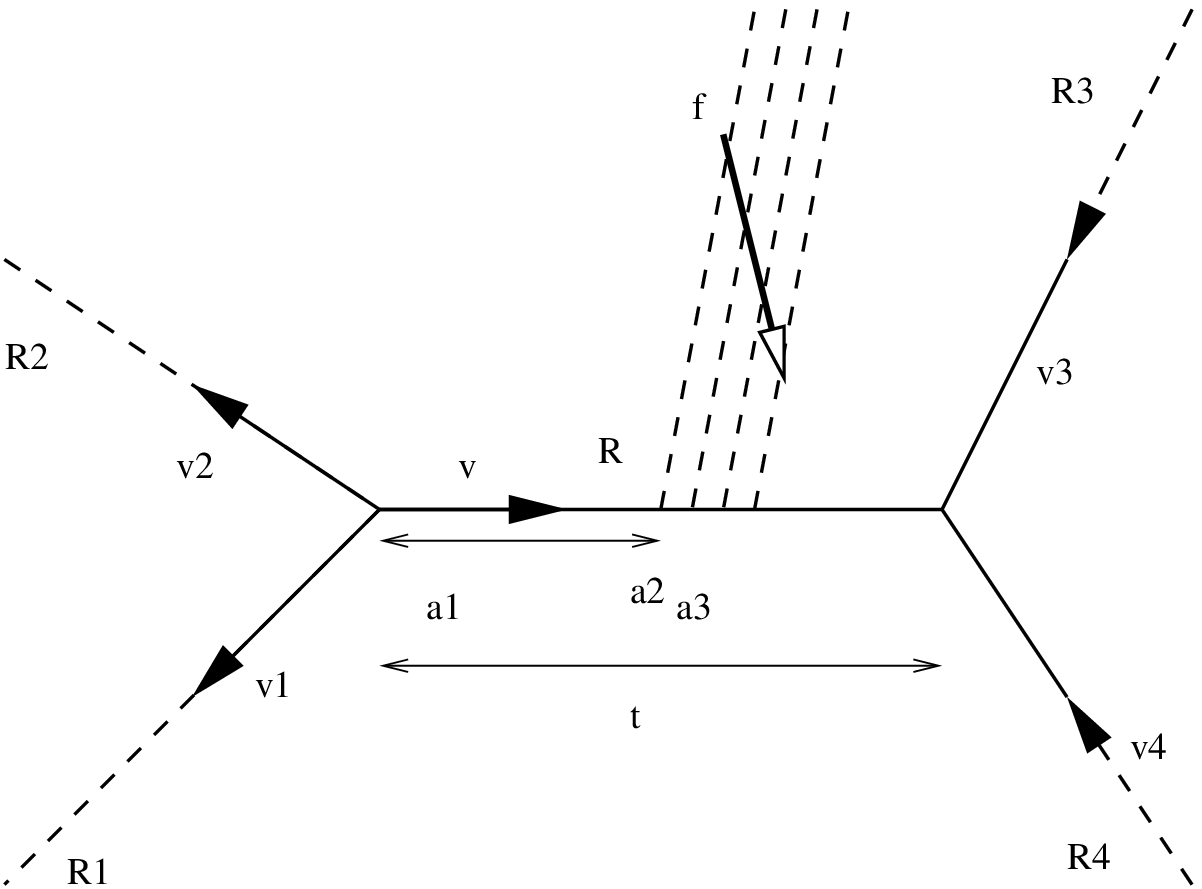}
&
&
\psfrag{R}{$R$}
\psfrag{R1}{$R_1$}
\psfrag{R2}{$R_2$}
\psfrag{R3}{$R_3$}
\psfrag{R4}{$R_4$}
\psfrag{R5}{$R_5$}
\psfrag{R6}{$R_6$}
\psfrag{a1}{$a$}
\psfrag{a2}{$g_sm$}
\psfrag{a3}{$t-a-g_sm$}
\psfrag{v}{$v$}
\psfrag{v1}{$v_1$}
\psfrag{v2}{$v_2$}
\psfrag{v3}{$v_3$}
\psfrag{v4}{$v_4$}
\psfrag{vinner}{
}
\psfrag{f}{$f$}
\includegraphics[scale=.55]{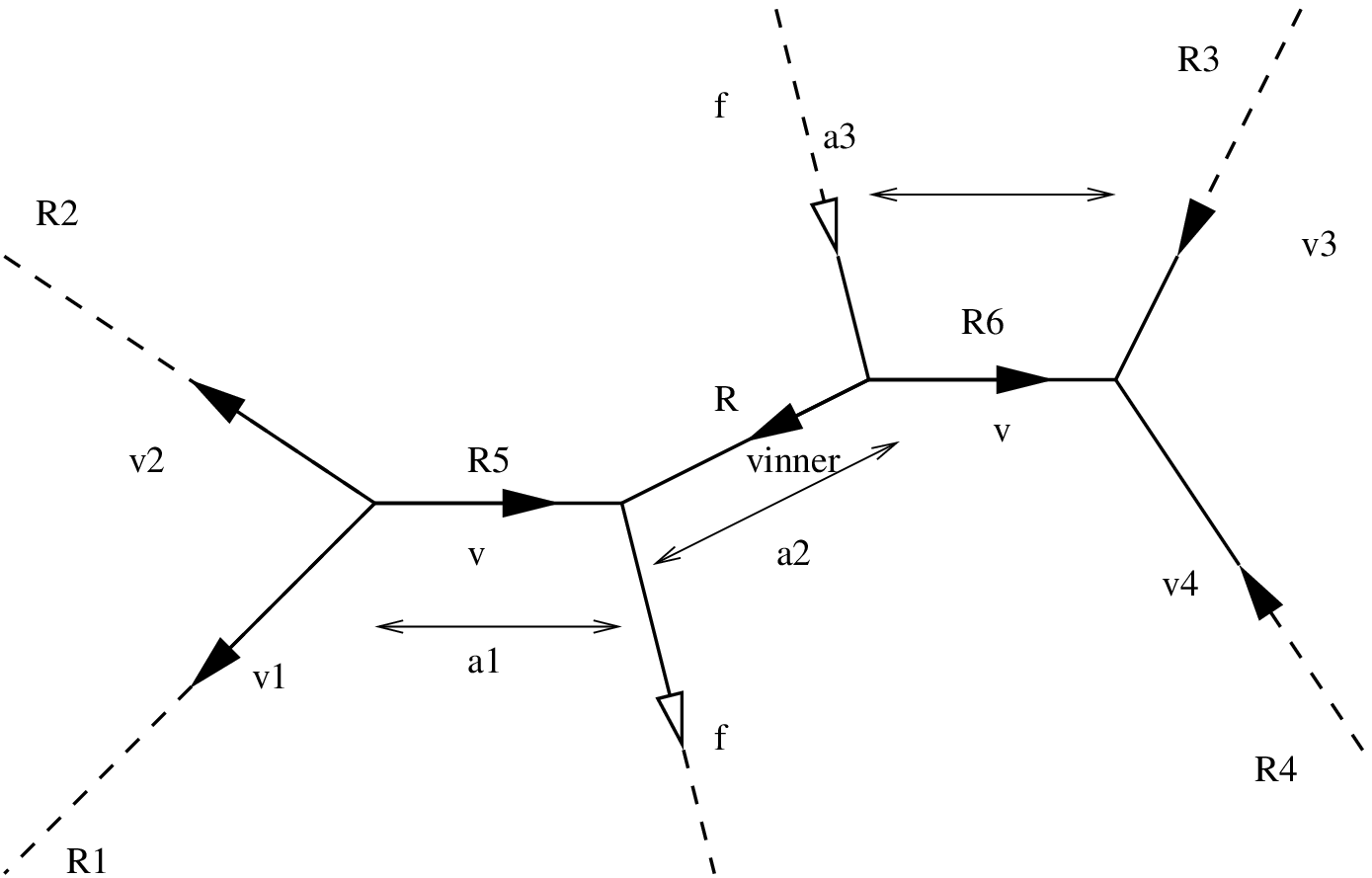}
\\
(a)&&(b)
\end{tabular}
\caption{
(a)
An internal edge of length $t$ in a toric web diagram.
$v,v_1,...,v_4$ are the vectors whose components are two coprime integers, and they specify the orientations
of the associated edges.
They satisfy the conditions $v_1\wedge v=v_2\wedge v_1=v\wedge v_2=1=v_3\wedge v=v_4\wedge v_3=v\wedge v_4$,
$v+v_1+v_2=0=v+v_3+v_4$.
$n:=v_1\wedge v_3$ is the relative framing of the two vertices.
We insert $m$ non-compact branes at the positions specified in the figure.
$f$ is another vector that specifies the framing of the branes, and satisfies the condition
 $f\wedge v=1$.
The integer $p:=f\wedge v_1$ enters the gluing rule of vertices.
(b)
After the geometric transition the branes get replaced by a new $S^2$ represented by
the edge of length $g_sm$.  The orientation of the new external edges is precisely given by the
framing vector of the branes.}
\label{vertex-1}
\end{figure}

Let us consider an arbitrary toric Calabi-Yau manifold that contains an interior
edge as shown in Figure \ref{vertex-1}(a).
Without $D$-branes the part of the partition function corresponding to this edge would be:
\ba
&&\sum_{R} C_{R_1R_2R}(-1)^{(n+1)|R|}
q^{\half n \kappa_R} e^{-|R|t}
C_{R^TR_3R_4}. \label{no-brane}
\ea
$t$ is the length of the edge, and $n$ is the relative framing of the two vertices. $C_{R_1R_2R_3}$ is the basic object underlying the topological vertex \cite{Aganagic:2003db}. $\kappa_R=|R|+\sum_iR_i^2-2iR_i$, where $R_i$ is the number of boxes on the $i$-th row and $|R|$ is the total number of boxes in the Young tableau $R$. See Appendix \ref{vertex-from-traces} for the explicit expression for $C_{R_1R_2R_3}$.

If we insert $D$-branes\footnote{
In the present convention, a brane here is an anti-brane in \cite{Gomis:2006mv} and vice versa.
This can be confirmed by computing a brane amplitude in the resolved conifold.
} with holonomy matrix $V$ in the internal edge,
(\ref{no-brane}) is replaced by:
\ba
&&\sum_{R,Q_L,Q_R} C_{R_1,R_2,R\otimes Q_L}(-1)^{s}
q^{-F} e^{-L}
C_{R^T\otimes Q_R,R_3,R_4}\Tr_{Q_L}V \Tr_{Q_R} V^\mo. \label{with-branes}
\ea
If the framing of the branes relative to the left vertex is $p$ then:\footnote{
Here $a=\int_D \Jcal$ is the complexified area of a holomorphic disk,
and $e^{-a}V$ is the gauge invariant open string modulus.
}
\ba
&&s=|R|+p(|R|+|Q_L|)+(n+p)(|R|+|Q_R|),\\
&&F=\half p\kappa_{R\otimes Q_L}+\half(n+p)\kappa_{R^T\otimes Q_R},~~
L=|R|t+|Q_L|a+|Q_R|(t-a).
\ea
Alternatively we can write (\ref{with-branes}) as:
\ba
&&\sum_{R_5,R_6} C_{R_1R_2R_5}\times(-1)^{p|R_5|}
q^{-\half p\kappa_{R_5}} e^{-|R_5|a} \left(\sum_R \Tr_{R_5/R}V  (-1)^{|R|}\Tr_{R_6/R^T} V^\mo\right)
\nn\\
&&
\times
(-1)^{(n+p)|R_6|}q^{-\half(n+p)\kappa_{R_6}}e^{-|R_6|(t-a)} C_{R_6R_3R_4}. \label{vertex-inner-brane-2}
\ea
Here $\Tr_{R/R'}(V):=\sum_{R''}N^{R}_{R'R''}\Tr_{R''}V$ 
 with $N^{R}_{R'R''}$
being 
tensor product coefficients.

In Appendix \ref{vertex-from-traces} we show that by substituting\footnote{
The exponent of $U_m$ differs from (\ref{holonomy-ground-state}) by an $i$-independent
shift that was absorbed in $a$.
}
\ba
V=U_{m}:={\rm
diag}(q^{m-i+1/2})_{i=1}^m,
\ea
that  the expression in the brackets in (\ref{vertex-inner-brane-2}),
multiplied by\footnote{
As we saw in section \ref{open2closedGV-section}, it is natural to include these factors
when considering branes with discrete values of the holonomy.
The product arises from annuli connecting the branes.
}
 $\xi(q)^m\prod_{1\leq i<j\leq m}(1-q^{j-i})$, is related to the topological vertex:
\ba
&&\xi(q)^{m}\prod_{1\leq i<j\leq m}(1-q^{j-i})\sum_R \Tr_{R_5/R}U_{m}  (-1)^{|R|}\Tr_{R_6/R^T} U_{m}^\mo
\nn\\
&=&
M(q)
q^{-m|R_6|}q^{-\half\kappa_{R_5}-\half \kappa_{R_6}}
\sum_R C_{\cdot R_5^TR} (-1)^{|R|}e^{-|R|g_sm} C_{R^T\cdot R_6^T}
 \label{vertex-identity}
\ea
The expression (\ref{vertex-inner-brane-2}) then becomes
\ba
&&M(q)\sum_{R,R_5,R_6} C_{R_1R_2R_5}(-1)^{p|R_5|}
q^{-\half (p+1)\kappa_{R_5}} e ^{-|R_5|a}
 C_{\cdot R_5^TR} (-1)^{|R|}e^{-|R|g_sm}
 \nn\\
&&\times
C_{R^T\cdot R_6^T}
(-1)^{(n+p+1)|R_6|}q^{-\half(n+p)\kappa_{R_6}}e^{-|R_6|(t-a-g_sm)} C_{R_6R_3R_4}. \label{vertex-geom-trans}
\ea
This is precisely the contribution from a part of the new geometry shown in Figure \ref{vertex-1}(b),
where the branes are replaced by a new $S^2$!
The orientations of the new edges are determined by the framing $p$ of the branes\footnote{
The equality of certain open and closed string amplitudes observed in section 3 of \cite{Caporaso:2006gk} 
is an example of the geometric transition discussed here.
We thank M. Marin\~o for pointing this out.
}.

\vskip+5pt
\noindent
{\it Anti-branes}
\vskip+5pt

We now demonstrate the geometric transition for  anti-branes.
Replacing branes by anti-branes is equivalent to  the replacement
$\Tr_R V \ra$ $ (-1)^{|R|}\Tr_{R^T}V$ \cite{Aganagic:2003db}.
Since $N^{R_1}_{R_2 R_3}=N^{R_1^T}_{R_2^T R_3^T}$\footnote{
This relation holds for $U(N)$ in the limit $N\ra \infty$,
and can be proven, for example, by using (\ref{charge-conjugation}) and (\ref{skewschur}).
},
this  is equivalent to replacing
the bracket in
(\ref{vertex-inner-brane-2}) by
$
(-1)^{|R_5|+|R_6|}$$\sum_R \Tr_{R_5^T/R}V$ $(-1)^{|R|}\Tr_{R_6^T/R^T} V^\mo .
$
Thus when anti-branes with holonomy $V$ are inserted,
the contribution from the part of geometry in Figure \ref{vertex-1}(a) is:
\ba
&&\sum_{R_5,R_6} C_{R_1R_2R_5}\times(-1)^{(p+1)|R_5|}
q^{-\half p\kappa_{R_5}} e^{-|R_5|a} \left(  \sum_R \Tr_{R_5^T/R}V  (-1)^{|R|}\Tr_{R_6^T/R^T} V^\mo \right)
\nn\\
&&\times
(-1)^{(n+p+1)|R_6|}q^{-\half(n+p)\kappa_{R_6}}e^{-|R_6|(t-a)} C_{R_6R_3R_4}. \label{vertex-inner-anti-brane}
\ea

\begin{figure}[ht]
\centering
\psfrag{R}{$R$}
\psfrag{R1}{$R_1$}
\psfrag{R2}{$R_2$}
\psfrag{R3}{$R_3$}
\psfrag{R4}{$R_4$}
\psfrag{R5}{$R_5$}
\psfrag{R6}{$R_6$}
\psfrag{a1}{$a$}
\psfrag{a2}{$g_sm$}
\psfrag{a3}{$t-a-g_sm$}
\psfrag{v}{$v$}
\psfrag{v1}{$v_1$}
\psfrag{v2}{$v_2$}
\psfrag{v3}{$v_3$}
\psfrag{v4}{$v_4$}
\psfrag{vinner}{
}
\psfrag{f}{$f$}
\includegraphics[scale=.55]{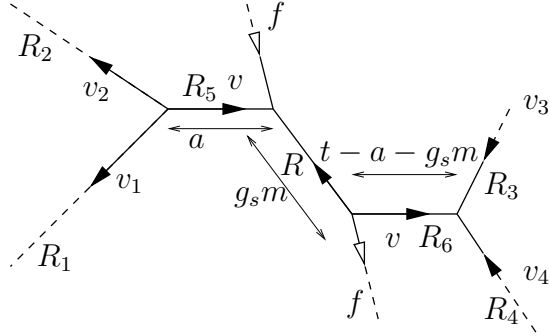}
\caption{The geometry that is obtained from Figure \ref{vertex-1}(a) through
geometric transition of anti-branes.
It is related to Figure \ref{vertex-1}(b) by flop.}
\label{vertex-2}
\end{figure}

Using the property that
$C_{R_1R_2R_3}=q^{-\half\kappa_{R_1}-\half \kappa_{R_2}-\half \kappa_{R_3}}C_{R_3^T R_2^T R_1^T}$
\cite{Aganagic:2003db}
we obtain from (\ref{vertex-identity}) the relation:
\ba
&&\xi(q)^{m}\prod_{1\leq i<j\leq m}(1-q^{j-i})(-1)^{|R_5|+|R_6|}
\sum_R \Tr_{R_5^T/R}U_{m}  (-1)^{|R|}\Tr_{R_6^T/R^T} U_{m}^\mo
\nn\\
&=&
M(q)
q^{-m|R_6|}
\sum_R C_{R_5^T\cdot R^T } (-1)^{|R|}e^{-|R|g_sm} C_{\cdot R R_6^T}.
 \label{vertex-identity-anti}
\ea
When combined with formula  (\ref{vertex-identity-anti}),
the amplitude (\ref{vertex-inner-anti-brane}) represents
the contribution from the part of the toric geometry shown in Figure \ref{vertex-2}.
This is related to the geometry in Figure \ref{vertex-1}(b) by a flop.
Again the orientations of the new edges are determined by the framing vector $f$ of the anti-branes.

\section*{Acknowledgments}
We are grateful to Mina Aganagic, Vincent Bouchard, Sergiy Koshkin,   Kentaro Hori and Marcos Mari\~no for useful discussions and correspondence.
We thank the Aspen Center for Physics where this project was
initiated. J.G. thanks l'\'Ecole Polytechnique for hospitality and   the European Union Excellence Grant MEXT-CT-2003-509661 for partial support. T.O. thanks the Perimeter Institute for Theoretical Physics for hospitality.
The research of T.O. is supported in part by the NSF grants PHY-05-51164 and PHY-04-56556.
Research at Perimeter Institute for Theoretical Physics is supported in
part by the Government of Canada through NSERC and by the Province of
Ontario through MRI. J.G. also acknowledges further support from an NSERC Discovery Grant.

 \appendix

 \renewcommand{\theequation}{\Alph{section}\mbox{.}\arabic{equation}}

 \bigskip\bigskip
 \noindent {\LARGE \bf Appendix}


\section{From open strings to closed strings}
\label{rewriteform}

In this Appendix we give a derivation of formula (\ref{cyl-amp}):
 \ba
&&\xi(q)^{P}\exp\left(-\sum_{n=1}^\infty\oo n\sum_{1\leq i<j\leq P} e^{-n(a_i-a_j)}\right) \nn\\
&=& M(q)^m\exp\left( \f{1}{ n\: [n]^{2}} \left[\sum_{1\leq I\leq J\leq 2m-1}  (-1)^{J-I+1}
e^{-n(t_I+t_{I+1}+...+t_J)}\right]\right).
\label{reproapp}
\ea
Using the value of the holonomies
\ba
a_i=g_s\left( R_i-i+P+\f{1}{ 2}\right), \qquad i=1,\ldots,P
\ea
the exponent on the left hand side of
(\ref{reproapp}) can be written as
\ba
S\equiv \sum_{n=1}^\infty\f{1}{ n}\sum_{1\leq i<j\leq P}e^{-ng_s(R_i-R_j)} e^{-ng_s(j-i)}.
\ea
We now perform the sum by using the parametrization of the Young tableau in Figure \ref{table}. There are two classes of contributions. The first class arises when   $(i,j)$  belong to the same ``block'' in the Young tableau, so that $R_i=R_j$, while the second class arises when  $(i,j)$ are in different ``blocks''  and $R_i\neq R_j$.
 The contribution from rows in the same ``block'' is given by
\ba
S_1=\sum_{n=1}^\infty\f{1}{ n}\sum_{I=1}^m \sum_{\alpha_I\leq i<j\leq \beta_I} e^{-ng_s(j-i)},
\label{sameblock}
\ea
while the contribution from rows in different ``blocks'' is
\ba
S_2=
 \sum_{n=1}^\infty\f{1}{ n}\sum_{1\leq I<J\leq m} e^{-ng_s\sum_{a=I}^{J-1}l_{2a}}\sum_{i=\alpha_I}^{\beta_I}e^{ng_s i}\sum_{j=\alpha_J}^{\beta_J}e^{-ng_s j},
 \label{secc}
\ea
where
\ba
\alpha_I=\sum_{a=1}^I l_{2a-3}+1 \qquad\hbox{and}\qquad  \beta_I=\sum_{a=1}^I l_{2a-1},
\ea
and $l_i$ $i=1,\ldots 2m+1$ are the coordinates of the Young tableau in Figure \ref{table}. In writing (\ref{secc}) we have used that the number of boxes in the $I$-th ``block'' is given by:
\ba
R_i=\sum_{a=I}^{m}l_{2a} \qquad i\in \hbox{{\it I}-th ``block''}.
\ea

The sum in (\ref{sameblock}) can be performed by grouping terms with the same value of $j-i$ and multiplying by the degeneracy; this yields:
\ba
\sum_{n=1}^\infty\f{1}{ n}\sum_{I=1}^m\sum_{k=1}^{\beta_I-\alpha_I}(\beta_I-\alpha_I+1-k)e^{-ng_sk}.
\ea
Using the formula
\ba
\sum_{k=1}^{c-1}(c-k)q^{nk}=-\f{1}{ \  [n]^2}\left[1-c(q^n-1)-q^{nc}\right]
\ea
we get that
\ba
S_1=-m\sum_{n=1}^\infty\f{1}{ n\: [n]^2}+P\sum_{n=1}^\infty \f{1}{n\: [n]}q^{n/2}+\sum_{I=1}^m\sum_{n=1}^\infty
\f{1}{ n\: [n]^2} e^{-nt_{2I-1}},
\label{s1final}
\ea
where $P=\sum_{I=1}^ml_{2I-1}$ is the number of rows in the Young tableau and $t_I=g_sl_I$.

The contribution from rows in different blocks can be straightforwardly computed using
\ba
\sum_{i=1+a}^b x^i=\f{x}{ 1-x}(x^a-x^b).
\ea
It is given by:
\ba
S_2=\sum_{n=1}^\infty \f{1}{ n\: [n]^2}\hskip-4pt \sum_{1\leq I<J\leq m}\hskip-4pt\left[e^{-n\sum_{a={2I}}^{2J-2}t_a}
\hskip-3pt+e^{-n\sum_{a={2I-1}}^{2J-1}t_a}\hskip-3pt-e^{-n\sum_{a={2I-1}}^{2J-2}t_a}\hskip-3pt
-e^{-n\sum_{a={2I}}^{2J-1}t_a}\right].
\label{s2final}
\ea
Therefore, combining (\ref{s1final}) and (\ref{s2final}) we get that
\ba
&&\xi(q)^{P}\exp\left(-\sum_{n=1}^\infty\oo n\sum_{1\leq i<j\leq P} e^{-n(a_i-a_j)}\right)=
M(q)^m\exp\Bigg(\sum_{n=1}^\infty -\f{1}{ n\: [n]^2}\Bigg[\sum_{I=1}^m e^{-nt_{2I-1}} \nn\\
&&+\sum_{1\leq I<J\leq m}\left[e^{-n\sum_{a={2I}}^{2J-2}t_a}+e^{-n\sum_{a={2I-1}}^{2J-1}t_a}-e^{-n\sum_{a={2I-1}}^{2J-2}t_a}-e^{-n\sum_{a={2I}}^{2J-1}t_a}\right]\Bigg]\Bigg),
\label{fincombine}
\ea
where
\ba
M(q)&\equiv &\exp\left(\sum_{n=1}^\infty \f{1}{n\: [n]^2}\right)=\prod_{n=1}^\infty\f{1}{(1-q^n)^n},\nn\\
\xi(q)&\equiv &\exp\left(\sum_{n=1}^\infty \f{1}{ n\: [n]^2}q^{n/2}\right)=\prod_{j=1}^\infty(1-q^j)^\mo.
\ea
The desired formula (\ref{reproapp}) then follows by combining the terms in (\ref{fincombine}).

Likewise, formula (\ref{tracerep})
\ba
\Tr_{\vec k} U_R^n=
\prod_{j=1}^\infty\left(\f{\sum_{I=1}^{m}   e^{-njT_{2I-1}}- e^{-njT_{2I}} }{ [nj]}\right)^{k_j}
\label{tracerepapp}
\ea
can be also be derived by splitting the sum over rows into blocks
\ba
\hbox{Tr}_{\vec k}U_R^n&=&
\prod_{j=1}^\infty \hskip-3pt\left(\sum_{i=1}^Pe^{-njg_s(R_i-i+P+1/2)}\hskip-3pt\right)^{k_j}\nn\\
&=&\prod_{j=1}^\infty \hskip-3pt\left(\sum_{I=1}^m
e^{-njg_s(\sum_{a=I}^ml_{2a}+\sum_{J=1}^ml_{2J-1})}
e^{-njg_s/2}\sum_{i=\alpha_I}^{\beta_I}e^{njg_s i}\hskip-3pt\right)^{k_j},
\ea
where we have used that $P=\sum_{J=1}^ml_{2J-1}$.
Now we can perform the sums to arrive at the right hand side of (\ref{tracerepapp}) by using that $T_I=\sum_{i=I}^{2m} g_s l_i$.

\section{Operator formalism}
\label{rewriteoperator}

In order to derive some of the group theory identities in the paper it is very convenient to exploit the relation between the representation theory of $U(N)$ and   two dimensional bosons and fermions in two dimensions.

Let us consider the mode expansion of a chiral boson $\phi(z)$ and fermions $\psi(z), \bar\psi(z)$ in two dimensions, which are
related by bosonization/fermionization:
\ba
&&\phi(z)=
i\sum_{n\neq 0} \f{\alpha_n}{nz^{n}},\\
&&\psi(z)=\sum_{r\in\Z+\half}\f{\psi_r}{z^{r+1/2}},
\bar\psi(z)=\sum_{r\in\Z+\half}\f{\bar\psi_r}{z^{r+1/2}},
\\
&&i\p\phi=:\psi\bar\psi:,~~\psi=:e^{i\phi}:,~~\bar\psi=:e^{-i\phi}:.
\label{modeexp}
\ea
The  oscillator modes satisfy the familiar commutation relations:
\ba
[\alpha_n,\alpha_m]=n\delta_{n+m,0},\qquad \{\psi_r,{\bar\psi_s}\}=\delta_{r+s,0}.
\ea
We can also define a charge conjugation operator $C$.
Charge conjugation $C$ exchanges $\psi$ and $\bar\psi$:
\ba
C\psi(z)C=\bar\psi(z),~C^2=1,~C|0\rangle=|0\rangle.
\ea
Then $C$ acts on $i\p\phi(z)=:\psi(z)\bar\psi(z):$ as:
\ba
C\p\phi(z)C=-\p\phi(z).
\ea
The connection between Young tableau $R$ and fermions arises from the identification
\ba
|R\rangle=\prod_{i=1}^d \psi_{-a_i-1/2}\bar\psi_{-b_i-1/2}|0\rangle,
\label{fermionstate}
\ea
where $a_i\equiv  R_i-i$,  $b_i=R_i^T-i$ are the Frobenius coodinates of $R$ and $d$ is the number of boxes in the diagonal of the Young tableau $R$.

It follows that:
\ba
C|R\rangle=(-1)^{|R|}|R^T\rangle. \label{charge-conjugation}
\ea

Let us now define \cite{Okounkov:2003sp} the operator
\ba
\Gamma_\pm(z):=\exp \sum_{n=1}^\infty \f{z^{\pm n}}{n} \alpha_{\pm
n},
\ea
which  satisfies
\ba
\Gamma_+(z_+)\Gamma_-(z_-)=\oo{1-z_+/z_-}\Gamma_-(z_-)\Gamma_+(z_+),~~
\Gamma_+(z)|0\rangle=|0\rangle,~\langle 0|\Gamma_-(z)=\langle 0|. \label{Gamma+-}
\ea
The skew Schur functions can be conveniently expressed as
\ba
s_{R/Q}(x)=\langle R|\prod_i\Gamma_-(x_i^\mo)|Q\rangle=\langle Q|\prod_i\Gamma_+(x_i)|R\rangle.
\label{skewschur}
\ea
The familiar Schur polynomials $s_R(x)$ arise when $|Q\rangle=|0\rangle$. In terms of these the skew Schur polynomials are given by
\ba
s_{R/Q}(x)=\sum_{R'}N^R_{QR'}s_{R'}(x),
\ea
where $N^R_{QR'}$ are the Littlewood-Richardson coefficients.

The following formula will come in handy in appendix \ref{vertex-from-traces}
\ba
e^{s L_0} \Gamma_\pm(z) e^{-sL_0}=\Gamma_\pm(e^{-s} z),
\ea
where
\ba
L_0=\sum_{n=1}^\infty\alpha_{-n}\alpha_n.
\ea

\section{An identity for integrality} \label{integralityapp}
Let us prove the equation (\ref{identity}).
\ba
\sum_{\vec k}\oo{z_{\vec k}}\chi_{R_1}(C(\vec k))
\prod_{j=1}^\infty \left(
\sum_{I=1}^{m}
\lambda_I{}^j-\sum_{I=1}^m \eta_I{}^j\right)^{k_j}=
\sum_{R_1,R_2,R_3}(-1)^{|R_3|}N^{R_1}_{R_2R_3} s_{R_2}(\lambda) s_{R_3^T}(\eta), ~~~~\label{identityapp}
\ea
This is a generalization of (7.29) in \cite{Aganagic:2002qg}.
It was used there for a similar purpose, and was originally derived
in \cite{Labastida:2000yw}.

\vskip+5pt
\noindent
{\it Proof}
\vskip+5pt

Consider oscillators $\alpha_n$ for a chiral boson as in (\ref{modeexp}). Let us consider the state
\ba
|\lambda,\eta\rangle&\equiv&
\sum_{\vec k}\oo{z_{\vec k}} \prod_{j=1}^\infty
 \left(\sum_{I=1}^{m}\lambda_I{}^j-\sum_{i=I}^m \eta_I{}^j
\right)^{k_j}\prod_{j=1}^\infty \alpha_{-j}^{k_j}|0\rangle
\nn\\
&=&\exp\left(
\sum_{n=1}^\infty \oo n\left(\sum_{I=1}^{m}\lambda_I{}^n-\sum_{I=1}^m\eta_I{}^n\right)
\alpha_{-n}\right)|0\rangle\nn\\
&=& \prod_{I=1}^{m}\Gamma_-(\lambda_I^\mo)\prod_{I=1}^m \Gamma_-^{-1}(\eta_I^\mo)|0\rangle.
\ea
The left hand side of (\ref{identityapp}) is $\langle
R_1|\lambda,\eta\rangle$, where $\langle R_1|$
is the fermionic Fock state associated with $R_1$ in (\ref{fermionstate}).

It can also be evaluated  as follows:
\ba
\langle R_1|\lambda,\eta\rangle&=&\sum_{R_2}\langle R_1|\prod_I
\Gamma_-(\lambda_I^\mo)|R_2\rangle\langle R_2| \prod_I \Gamma_-^\mo(\eta_I^\mo)|0\rangle\nn\\
&=&\sum_{R_2}\langle R_1|\prod_I
\Gamma_-(\lambda_I^\mo)|R_2\rangle(-1)^{|R_2|}\langle R_2^T| C\prod_I \Gamma_-^\mo(\eta_I^\mo)|0\rangle\nn\\
&=&\sum_{R_2}(-1)^{|R_2|} s_{R_1/R_2}(\lambda) s_{R_2^T}(\eta)\nn\\
&=&\sum_{R_2,R_3}(-1)^{|R_2|}N^{R_1}_{R_2R_3} s_{R_3}(\lambda)s_{R_2^T}(\eta).
\ea
where we have used (\ref{skewschur}). This proves (\ref{identityapp})
after relabeling $R_2$ and $R_3$.

\section{From closed strings to open strings} \label{closed2open-appendix}

On the right hand side of (\ref{open2closedGV}), let us focus on:
\ba
\sum_{R_1R_2R_3}
 {\hat N}_{R_1g{\vec Q}} (-1)^{|R_3|}N^{R_1}_{R_2R_3}
s_{R_2}(e^{-T_o}) s_{R_3^T}(e^{-T_e}). \label{eqn-rep-basis}
\ea
In the operator formalism, we can write this as:
\ba
\sum_{R_1} {\hat N}_{R_1g{\vec Q}} \langle R_1| \prod_i \Gamma_-(e^{T_{o,i}})
 \prod_i \Gamma_-(e^{T_{e,i}})^\mo|0\rangle.
\ea
If we define
\ba
|\vec k\rangle=\prod_{j=1}^\infty \alpha_{-j}^{k_j}|0\rangle,
~~~N_{\vec k g {\vec Q}}:=\sum_{R_1}\chi_{R_1}(C(\vec k)) {\hat N}_{R_1g{\vec Q}},
\ea
then
\ba
|R\rangle=\sum_{\vec k}\oo{z_{\vec k}}\chi_R(C(\vec k))|\vec k\rangle.
\ea
Thus (\ref{eqn-rep-basis}) equals
\ba
\sum_{\vec k}\oo{z_{\vec k}}{\hat N}_{\vec k g{\vec Q}}
\langle \vec k| \prod_i \Gamma_-(e^{T_{o,i}})
 \prod_i \Gamma_-(e^{T_{e,i}})^\mo|0\rangle.
\ea
Notice that
\ba
\prod_i \Gamma_-(x_i^\mo)^\pm&=&\exp \pm \sum_{j=1}^\infty \oo j \sum_i x_i^j \alpha_{-j}\nn\\
&=&\sum_{\vec k}\f{(\pm 1)^{\sum_j k_j}}{z_{\vec k}} P_{\vec k}(x)\prod_{j=1}^\infty \alpha_{-j}^{k_j}.
\ea
Here $P_{\vec{k}} (x)=\prod_j (\sum_i x_i^j)^{k_j}$.
(\ref{eqn-rep-basis}) becomes
\ba
&&\sum_{\vec k}\oo{z_{\vec k}}{\hat N}_{\vec k g{\vec Q}}
\langle \vec k|\sum_{\vec k_1, \vec k_2}\f{(- 1)^{\sum_j k_{2,j}}}{z_{\vec k_1}z_{\vec k_2}}
P_{\vec k_1}(e^{-T_o})P_{\vec k_2}(e^{-T_e})|\vec k_1+\vec k_2\rangle\nn\\
&=&\sum_{\vec k_1, \vec k_2}{\hat N}_{\vec k_1+\vec k_2, g,{\vec Q}}
\f{(- 1)^{\sum_j k_{2,j}}}{z_{\vec k_1}z_{\vec k_2}}
P_{\vec k_1}(e^{-T_o})P_{\vec k_2}(e^{-T_e}).\label{eqn-winding-basis}
\ea

It is clear that the contributions from (\ref{eqn-rep-basis}) to
\ba
 \sum_{\vec Q_b}n_g^{\vec Q_b}(X_b)
e^{-\vec Q_b\cdot\vec t}
\ea
have to be symmetric with respect to $e^{-T_{o,i}}$, and also with respect to $e^{-T_{e,i}}$.
Any such function can be expanded in $P_{\vec k_1}(e^{-T_o})P_{\vec k_2}(e^{-T_e})$,
and once we know the coefficients, we can read off ${\hat N}_{\vec k g{\vec Q}}$ using (\ref{eqn-winding-basis}).
Finally one computes the open GV invariants using the formula
${\hat N}_{R  g{\vec Q}}=\sum_{\vec k} {\hat N}_{\vec k g{\vec Q}}\chi_R(C(\vec k))/z_{\vec k}$.

\section{Topological vertex amplitude}\label{top-vert-app}
We use the convention such that $q$ is replaced by $q^\mo$ relative to \cite{Aganagic:2003db}.
Explicitly it is given, with slight abuse of notation, by:
\ba
\hspace{-2mm}
C_{R_1R_2R_3}(q)=q^{-\half(\kappa_{R_2}+\kappa_{R_3})}
s_{R_2^T}(q^{i-1/2})\sum_Q s_{R_1/Q}(q^{-(R_2^T)_i+i-1/2})
s_{R_3^T/Q}(q^{-(R_2)_i+i-1/2}).
\ea
Here $s_{R_1/R_2}$ is a skew Schur function.  The index $i$ runs from $1$ to $\infty$.


The partition function of topological strings on any toric Calabi-Yau manifold,
with or without $D$-branes, can be computed by gluing several topological vertices.
The gluing rules are explained in subsection \ref{geom-trans-toric}.

\section{An identity for geometric transitions}\label{vertex-from-traces}
In this appendix we prove the identity (\ref{vertex-identity}).

First we compute
\ba
&&\sum_R \Tr_{R_5/R}U_{m\times l}   (-1)^{|R|}\Tr_{R_6/R^T} U_{m\times l} ^\mo\nn\\
&=&\sum_R \langle R_5|\prod_{i=1}^m\Gamma_-(q^{-l-m+i-1/2})|
R\rangle (-1)^{|R|}\langle
R^T|\prod_{i=1}^m\Gamma_+(q^{-l-m+i-1/2})|R_6\rangle
\nn\\
&=&\sum_R \langle R_5|\prod_{i=1}^m\Gamma_-(q^{-l-m+i-1/2})|
R\rangle \langle
R|C\prod_{i=1}^m\Gamma_+(q^{-l-m+i-1/2})|R_6\rangle
\nn\\
&=&(-1)^{|R_6|} \langle R_5|\prod_{i=1}^m\Gamma_-(q^{-l-m+i-1/2})
\prod_{i=1}^m\Gamma_+^\mo(q^{-l-m+i-1/2})|R_6^T\rangle
\nn\\
&=&(-1)^{|R_6|} \langle R_5|\prod_{i=1}^\infty\Gamma_-(q^{-l-i+1/2})
\prod_{i=1}^\infty\Gamma_-^\mo(q^{-l-m-i+1/2})
\nn\\
&&\times\prod_{i=1}^\infty\Gamma_+^\mo(q^{-l-m+i-1/2})
\prod_{i=1}^\infty\Gamma_+(q^{-l+i-1/2})
|R_6^T\rangle\nn\\
&=&(-1)^{|R_6|} \prod_{i,j=1}^\infty (1-q^{m+i+j-1})^\mo
 \langle R_5|\prod_{i=1}^\infty\Gamma_-(q^{-l-i+1/2})
\prod_{i=1}^\infty\Gamma_+(q^{-l+i-1/2})
\nn\\
&&\times
\prod_{i=1}^\infty\Gamma_-^\mo(q^{-l-m-i+1/2})
\prod_{i=1}^\infty\Gamma_+^\mo(q^{-l-m+i-1/2})
|R_6^T\rangle
\nn\\
&=& (-1)^{|R_6|} e^{\sum_{n=1}^\infty \f{e^{-n g_sm}}{n[n]^2}}
 \sum_{R,Q,Q'}
 \langle R_5|\prod_{i=1}^\infty\Gamma_-(q^{-l-i+1/2})|Q\rangle \langle Q|
\prod_{i=1}^\infty\Gamma_+(q^{-l+i-1/2}) |R\rangle\nn\\
&&\times(-1)^{|R|}\langle R^T|
\prod_{i=1}^\infty\Gamma_-(q^{-l-m-i+1/2}) |Q'\rangle \langle Q'|
\prod_{i=1}^\infty\Gamma_+(q^{-l-m+i-1/2})
|R_6\rangle(-1)^{|R_6|}
\nn\\
&=&  e^{\sum_{n=1}^\infty \f{e^{-n g_sm}}{n[n]^2}}
 \sum_{R,Q,Q'}(-1)^{|R|} q^{l(|R_5|-|R|)}
 s_{R_5/Q}(q^{i-1/2})
 s_{R/Q}(q^{i-1/2})
\nn\\
&&\times q^{(l+m)(|R|-|R_6|)}
s_{R^T/Q'}(q^{i-1/2}) s_{R_6/Q'}(q^{i-1/2})
\nn\\
&=&  q^{l|R_5|-(l+m)|R_6|-\half\kappa_{R_5}-\half \kappa_{R_6}}
e^{\sum_{n=1}^\infty \f{e^{-n g_sm}}{n[n]^2}}
 \sum_{R}
 C_{\cdot R_5^T R} (-1)^{|R|}e^{-|R|g_sm}C_{R^T\cdot R_6^T}.
\ea
Combining this with (\ref{cyl-amp}) when $R_i=0$ gives (\ref{vertex-identity}).

\bibliography{geom-trans19}

\end{document}